\documentclass[aps,twocolumn,amsmath,amssymb,showpacs]{revtex4-1}
\usepackage{epsfig,color,graphicx,ulem,float}
\usepackage{xcolor}
\begin{document}
\def\be{\begin{equation}}
\def\ee{\end{equation}}
\def\bea{\begin{eqnarray}}
\def\eea{\end{eqnarray}}

\def\a{\alpha}
\def\b{\beta}
\def\g{\gamma}
\def\d{\delta}
\def\e{\eta}
\def\r{\rho}
\def\th{\theta}
\def\ph{\phi}
\def\eps{\epsilon}

\newenvironment{ale}{\color{blue}}

\def\fr{\frac}
\def\l{\label}
\def\th{\theta}
\newcommand{\dd}{\mbox{d}}
\newcommand{\gae}{\lower 2pt \hbox{$\, \buildrel {\scriptstyle >}\over {\scriptstyle
\sim}\,$}}
\newcommand{\lae}{\lower 2pt \hbox{$\, \buildrel {\scriptstyle <}\over {\scriptstyle
\sim}\,$}}

\title{Nonequilibrium first-order phase transition in coupled oscillator
systems with inertia and noise}
\author{Shamik Gupta$^1$, Alessandro Campa$^2$ and Stefano Ruffo$^3$}
\affiliation{$^1$Laboratoire de Physique Th\'{e}orique et Mod\`{e}les
Statistiques (UMR CNRS 8626), Universit\'{e} Paris-Sud, Orsay, France\\
$^{2}$Health and Technology Department, Istituto Superiore di Sanit\`{a},
and INFN Sezione Roma1, Gruppo Collegato Sanit\`{a}, Roma, Italy \\
$^3$Department of Physics and Astronomy and CSDC, University of Florence, CNISM
and INFN,  via G. Sansone, 1 50019 Sesto Fiorentino, Italy}
\begin{abstract}
We study the dynamics of a system of coupled oscillators of distributed natural
frequencies, by including the features of both thermal
 noise, parametrized by a temperature, and inertial terms,  parametrized by a moment of inertia.
  For a general unimodal frequency distribution, we
report here the complete phase diagram of the model
in the space of dimensionless moment of inertia, temperature, and width
of the frequency distribution.
 We demonstrate that the system
undergoes a nonequilibrium first-order phase transition from a
synchronized phase at low parameter values to an incoherent phase at high values. 
 We provide strong numerical evidence for the existence
of both the synchronized and the incoherent phase, treating the latter analytically to
obtain the corresponding linear stability threshold that bounds the
first-order transition point from below.
In the limit of zero noise and inertia, when the dynamics reduces to the one of
the Kuramoto model, we recover the associated known continuous
transition. At finite noise and inertia but in
the absence of natural frequencies, the
dynamics becomes that of a well-studied model of long-range
interactions, the Hamiltonian mean-field model.
Close to the first-order phase transition, we show that the escape time out of
metastable states scales exponentially with the number of oscillators,
which we explain to be stemming from the long-range nature of the
interaction between the oscillators.
\end{abstract} 
\pacs{05.70.Fh, 05.70.Ln, 05.45.Xt}
\maketitle
\section{Introduction}
Collective synchronization refers to the remarkable phenomenon
of a large population of
coupled oscillators spontaneously synchronizing to oscillate at a
common frequency, despite each constituent having a different
natural frequency.
This many-body cooperative effect is
observed in many physical and biological systems, pervading length
 and time scales of several orders of magnitude. Some examples are
metabolic synchrony in yeast cell suspensions \cite{Bier:2000}, synchronized
firings of cardiac pacemaker cells \cite{Winfree:1980}, flashing in
unison by groups of fireflies \cite{Buck:1988}, 
voltage oscillations at a common frequency in an array of
current-biased Josephson junctions \cite{Wiesenfeld:1998},
phase synchronization in electrical power distribution networks
\cite{Filatrella:2008,Rohden:2012,Dorfler:2013}, rhythmic applause
\cite{Neda:2000}, animal
flocking behavior \cite{Ha:2010}; see Ref. \cite{Strogatz:2003} for a recent survey.

A paradigmatic model to study synchronization is the Kuramoto model
comprising $N$ phase-only oscillators of distributed natural 
frequencies that are globally coupled through the sine of their phase differences
\cite{Kuramoto:1984,Strogatz:2000}.
Specifically, the system involves $N$ interacting oscillators
$i=1,2,\ldots,N$. The $i$-th oscillator has natural frequency
$\omega_i$, and is
characterized by its phase $\th_i$ which is a
periodic variable of period $2\pi$. The $\omega_i$'s have a common
probability distribution given by $g(\omega)$. The phase $\th_i$
evolves in time according to the equation
\be
\fr{\dd\th_i}{\dd t}=\omega_i+\fr{\widetilde{K}}{N}\sum_{j=1}^N\sin(\th_j-\th_i),
\l{timeevolution}
\ee
where $\widetilde{K}$ is the coupling constant, while the factor $1/N$ makes the
model well behaved in the continuum limit $N \to \infty$.
 
In this work, we study a
generalization of the dynamics Eq. (\ref{timeevolution}) that includes inertial terms parametrized by
a moment of inertia and stochastic noise parametrized by a temperature
\cite{Acebron:1998,Hong:1999,Acebron:2000}.
Noise accounts for the temporal fluctuations of the natural frequencies \cite{Sakaguchi:1988}, while
inertia elevates the first-order Kuramoto dynamics to second-order
\cite{Ermentrout:1991}. For a general unimodal distribution of the
natural frequencies, we
report here the complete phase diagram of the model
in the space of dimensionless moment of inertia, temperature, and width
of the frequency distribution, showing that the system in the steady state
may exist in either of two possible phases, namely, a synchronized phase
and an unsynchronized or incoherent phase. We show that a
nonequilibrium first-order transition occurs from the synchronized phase at low
parameter values to the incoherent phase at high values.
While strong
numerical evidence is provided to support the existence of both the
synchronized and the incoherent phase, only the latter could be treated analytically to
obtain the corresponding linear stability threshold that bounds the
first-order transition point from below. In proper
limits of the dynamics, we recover the known continuous phase transitions in the Kuramoto
model and in its noisy extension \cite{Sakaguchi:1988}, and an equilibrium continuous
transition in a related model of long-range interactions,
the Hamiltonian mean-field model \cite{Inagaki:1993}. 

The Kuramoto model has been almost exclusively studied within the field of
synchronization and non-linear dynamical systems. On the other hand,
there has been much recent activity within the community of
statistical physicists to study nonequilibrium stationary states (NESSs) and 
develop a general framework akin to the one due to Boltzmann and Gibbs for
equilibrium that allows analysis of nonequilibrium states on a general
footing \cite{Mallick:2009}. Unfortunately, there are few
examples of NESSs for which one knows the probability measure of
configurations exactly, so that the bulk of studies have
relied on numerical simulations and approximate analysis
\cite{Privman:1997}.

Our work interprets the dynamics of the Kuramoto model to be of
true non-equilibrium character. Moreover, quenched disorder in the form
of natural frequencies of the oscillators provides a very rich setting
to study the interplay of the nonequilibrium character of the dynamics
with the disorder. In this rich backdrop, we are able to characterize
the nature of the NESS and ascertain under quite general conditions the whole spectrum of phase transitions.

The paper is organized as follows. In the following section, we describe
the model of interest and briefly review previous studies of the model. In
Sec. \ref{phase-diagram}, we present the complete phase diagram of
the model, providing numerical simulation results in support. In Sec. \ref{analytical},
we present an analytical treatment of the properties of the incoherent phase, based on
the Kramers equation for the single-oscillator distribution. This is
followed in Sec. \ref{numerical} by a comparison of our analytical
predictions with numerical simulations. The paper ends with conclusions.
Some of the technical details are relegated to the two appendices.

\section{The model}
\l{model}
We now give a precise definition of the generalized dynamics that we study in
this paper. In addition to phase $\th_i$, we associate with the $i$-th
oscillator another dynamical variable, namely, the
 angular velocity $v_i$. With a Gaussian noise force
$\eta_i(t)$ and the natural frequency $\omega_i$, the dynamics is \cite{Acebron:1998,Acebron:2000} 
\be
\frac{\dd\theta_i}{\dd t}=v_i, m\frac{\dd v_i}{\dd t}=-\gamma v_i+K
r\sin(\psi-\th_i)+\gamma \omega_i+\sqrt{\gamma}\eta_i(t),
\l{eom}
\ee
where $m$ is the oscillator moment of inertia, $\gamma$ is the friction
constant, while $r$ is the synchronization order parameter:
\be
r(t)e^{i\psi(t)} \equiv \fr{\sum_{j=1}^N e^{i\th_j(t)}}{N}.
\ee
Here, we have 
\be
\langle \eta_i(t) \rangle=0, \langle \eta_i(t)\eta_j(t')
\rangle=2T\delta_{ij}\delta(t-t'),
\ee
with temperature $T$ in units of the
Boltzmann constant. We consider a unimodal
$g(\omega)$ (that is, symmetric about mean $\widetilde{\omega}$, and
decreases to zero with increasing $|\omega-\widetilde{\omega}|$), and
denote its width by $\sigma$. In the absence
of inertia, the dynamics with the redefinition $K/\gamma=\widetilde{K}$ reduces at $T=0$ to that of the Kuramoto model
\cite{Kuramoto:1984,Strogatz:2000} and at $T \ne 0$ to that of its
extension studied by Sakaguchi in Ref. \cite{Sakaguchi:1988}. 

The dynamics Eq. (\ref{eom}) also describes motion of particles with an $XY$-interaction on a unit circle, with
$\th_i, v_i$ and $\gamma \omega_i$ being respectively the angular coordinate,
velocity and external torque. In the absence of $\omega_i$'s, Eq. (\ref{eom}) for $\gamma=0$ is
the microcanonical dynamics of the Hamiltonian mean-field model \cite{Inagaki:1993}, a
prototype of long-range interacting systems \cite{Campa:2009}. In this
case, the equations of motion are the Hamilton equations associated
with the Hamiltonian
\be
H=\sum_{i=1}^N
\frac{p_i^2}{2m}+\frac{K}{2N}\sum_{i,j=1}^N\Big[1-\cos(\th_i-\th_j)\Big],
\l{H}
\ee
with $p_i=mv_i$ the momentum of the $i$-th particle.
The dynamics of this system is microcanonical, conserving energy and
total momentum. With no $\omega_i$'s, but $\gamma \ne 0$, the dynamics of the resulting Brownian mean-field (BMF) model is canonical, mimicking
the interaction of the HMF system with a heat bath \cite{Chavanis:2013}.

The dynamics Eq. (\ref{eom}) is invariant under $\th_i
\to \th_i+\widetilde{\omega}t, v_i \to
v_i+\widetilde{\omega}, \omega_i \to
\omega_i+\widetilde{\omega}$, and the effect of $\sigma$ may be made explicit by replacing 
$\omega_i$ in the second equation with $\sigma \omega_i$.
We thus consider from now on the dynamics Eq. (\ref{eom}) with
the substitution $\omega_i \rightarrow \sigma \omega_i$. In the resulting
model, we take $g(\omega)$ to have zero mean and unit width, without loss of 
generality.

For $m \ne 0$, using dimensionless variables 
\bea
&&\overline{t}\equiv 
t\sqrt{K/m}, \\
&&\overline{v}_i\equiv
v_i\sqrt{m/K}, \\ 
&&1/\sqrt{\overline{m}}\equiv
\gamma/\sqrt{Km}, \\
&&\overline{\sigma} \equiv \gamma \sigma/K,\\
&&\overline{T} \equiv
T/K, \\
&&\overline{\eta}_i(\overline{t})\equiv
\eta_i(t)\sqrt{\gamma}/K,
\eea
the dynamics becomes
\bea
\frac{\dd\th_i}{\dd\overline{t}}=\overline{v}_i,\frac{\dd\overline{v}_i}{\dd\overline{t}}=
-\frac{1}{\sqrt{\overline{m}}}\overline{v}_i+r\sin(\psi-\th_i)+\overline{\sigma}\omega_i+\overline{\eta}_i(\overline{t}),
\nonumber \\
\l{eom-scaled} 
\eea
where 
\be
\langle \overline{\eta}_i(\overline{t})\overline{\eta}_j(\overline{t}')
\rangle=2\fr{\overline{T}}{\sqrt{\overline{m}}}\delta_{ij}\delta(\overline{t}-\overline{t}').
\ee
For $m=0$, using dimensionless time $\overline{t}\equiv
t(K/\gamma)$, the dynamics becomes the overdamped motion
\be
\fr{\dd\th_i}{\dd\overline{t}}=r\sin(\psi-\th_i)+\overline{\sigma}\omega_i+\overline{\eta}_i(\overline{t}),
\l{overdamped}
\ee
where 
\be
\langle \overline{\eta}_i(\overline{t})\overline{\eta}_j(\overline{t}')
\rangle=2\overline{T}\delta_{ij}\delta(\overline{t}-\overline{t}').
\ee
From now on, we will consider in place of dynamics Eq. (\ref{eom}) the
reduced dynamics Eq. (\ref{eom-scaled}) [that reduces for $m=0$ to the
overdamped dynamics Eq. (\ref{overdamped})] involving three
dimensionless parameters, $\overline{m},\overline{T},\overline{\sigma}$;
we will drop overbars for simplicity of notation.
With $\sigma=0$ (i.e. $g(\omega)=\delta(\omega)$ 
\cite{Acebron:1998},\cite{Acebron:2000}),
the resulting BMF dynamics has an equilibrium stationary
state \cite{Chavanis:2013}. For other $g(\omega)$, the
dynamics Eq. (\ref{eom-scaled}) violates detailed balance due to the
external driving by the set of torques $\{\gamma
\omega_i\}$, yielding a NESS. We demonstrate this in 
Appendix \ref{app0}.

Several stationary state aspects of the dynamics Eq. (\ref{eom-scaled}) in
the continuum limit $N \to \infty$ are known. For the Kuramoto
dynamics ($m=T=0$), the system exhibits a continuous transition from a
low-$\sigma$ synchronized [$r_{\rm st}=r(t \to \infty) \ne 0$] to a high-$\sigma$ incoherent
($r_{\rm st}=0$) phase across the critical point \cite{Kuramoto:1984}
\be
\sigma_c(m=0,T=0)=\fr{\pi g(0)}{2};
\l{kura-result}
\ee
extending to $T \ne 0$, the point becomes a
second-order critical line $\sigma_c(m=0,T)$ on the $(T,\sigma)$-plane, given, on using
the results of Sakaguchi in Ref.
\cite{Sakaguchi:1988}, by solving 
\be
2=\int_{-\infty}^\infty \fr{Tg(\omega)\dd \omega}{T^2+\omega^2\sigma^2_c(m=0,T)}.
\l{saka-result}
\ee
For the BMF dynamics ($\sigma=0; ~m,T \ne 0$), the synchronization transition
is again continuous, occurring at the critical temperature given by \cite{Chavanis:2013} 
\be
T_c=\fr{1}{2}.
\l{bmf-tc}
\ee
Although there have been some numerical studies of
the full dynamics for non-zero $m, T, \sigma$ \cite{Tanaka:1997,Acebron:1998,Acebron:2000},  the complete
synchronization phase diagram for a general unimodal $g(\omega)$ has not
been addressed before, a question we take up and answer in this paper.
In the next section, we describe the complete phase diagram that emerges
out of our analysis.

\begin{figure}[!h]
\centering
\includegraphics[width=105mm]{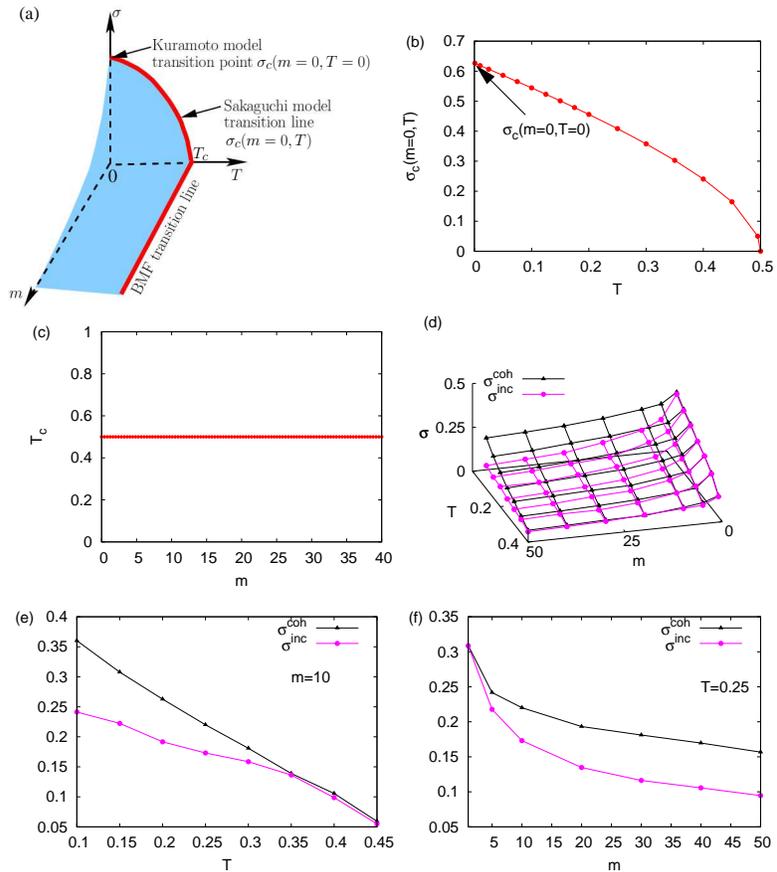}
\caption{(Color online) (a) Schematic phase diagram of model Eq. 
(\ref{eom-scaled}) in terms of
dimensionless moment of inertia $m$, temperature $T$, and width $\sigma$ of the
frequency distribution: the shaded blue
surface is a first-order transition surface, the thick red lines are
second-order critical lines. The system is synchronized inside the region
bounded by the surface, and is incoherent outside. The limits (the Kuramoto model, the Sakaguchi model
and the BMF
model) in which
known transitions are obtained are labeled. (b) The known transition line for the Sakaguchi
model, given by Eq. (\ref{saka-result}), showing also the Kuramoto model
transition point, Eq. (\ref{kura-result}), for a Gaussian $g(\omega)$
with zero mean and unit width \cite{width}. (c) The known
transition line for the BMF model, given by Eq. (\ref{bmf-tc}). The
shaded blue surface in (a) is bounded
from above and below by the dynamical stability thresholds $\sigma^{\rm coh}(m,T)$
and $\sigma^{\rm inc}(m,T)$ of the synchronized and the incoherent phase
respectively. These thresholds may be estimated in $N$-body
simulations from hysteresis plots (see Fig. \ref{fig2} for an example);
Panel (d) shows the surfaces $\sigma^{\rm coh}(m,T)$
and $\sigma^{\rm inc}(m,T)$ obtained from $N$-body simulations with
$N=500$ for a Gaussian $g(\omega)$ with zero mean and unit width, with
cuts of the three-dimensional plot at $m=10$ shown in panel (e) and at
$T=0.25$ shown in panel (f).}
\l{fig1}
\end{figure}

\begin{figure}[here!]
\centering
\includegraphics[width=80mm]{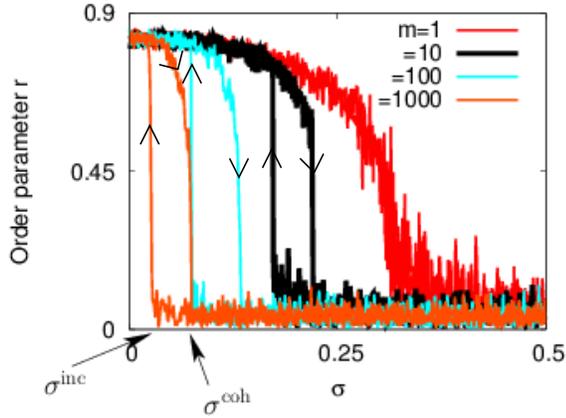}
\caption{(Color online) $r$ vs. adiabatically tuned
$\sigma$ for different $m$ values at $T=0.25<T_c=1/2$, showing also the stability thresholds, $\sigma^{\rm inc}(m,T)$ and
$\sigma^{\rm coh}(m,T)$, for $m=1000$. For a given $m$, the branch of the plot to
the right (left) marked with an arrowhead pointing down 
(up) corresponds to $\sigma$ increasing (decreasing); for
$m=1$, the two branches almost overlap. The data are obtained in
$N$-body simulations with $N=500$ for a
Gaussian $g(\omega)$ with zero mean and unit width.}
\l{fig2}
\end{figure}

\begin{figure}[here]
\centering
\includegraphics[width=80mm]{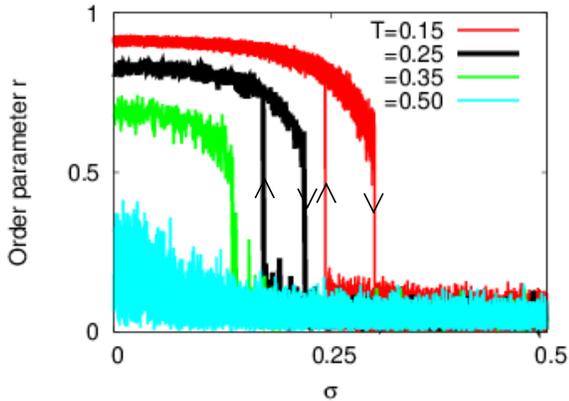}
\caption{(Color online) $r$ vs. adiabatically tuned
$\sigma$ for different temperatures $T \le T_c=1/2$ at a fixed moment of
inertia $m=10$. 
For a given $T$, the branch of the plot to
the right (left) marked with an arrowhead pointing down 
(up) corresponds to $\sigma$ increasing (decreasing); for $T \ge 0.35$,
the two branches almost overlap. The data are obtained in
$N$-body simulations with $N=500$ for a
Gaussian $g(\omega)$ with zero mean and unit width. Similar
disappearance of the hysteresis loop with increase of $T$ was reported in Ref.
\cite{Hong:1999}.}
\l{fig3}
\end{figure}

\section{Phase diagram}
\l{phase-diagram}
The complete phase diagram is shown schematically in Fig. \ref{fig1}(a), where the
thick red second-order critical lines stand for the continuous transitions mentioned above. For non-zero
$m,T,\sigma$, the synchronization transition becomes first-order,
occurring across the shaded blue transition surface; this surface is bounded by the
second-order critical lines on the $(T,\sigma)$ and $(m,T)$ planes, and by a first-order transition line on
the $(m,\sigma)$-plane. 

The phase diagram in Fig. \ref{fig1}(a) is a
generalization of the one for typical fluids where a first-order
transition line ends in a critical point, while we have here a first-order
transition surface ending in critical lines.
All transitions for $\sigma \ne 0$ are in NESS, and we
interpret them to be of dynamical origin, accounted for by stability
considerations of stationary solutions of equations describing evolution
of phase-space distribution. Showing that the phases extremize a
free-energy-like quantity (e.g., a large deviation functional
\cite{Touchette:2009}) in NESS is a daunting task in
the absence of a general framework akin to that for equilibrium \cite{Derrida:2005}. For $\sigma=0$, the
different phases actually minimize the equilibrium free energy \cite{Campa:2009}. 
\begin{figure}[here!]
\centering
\includegraphics[width=80mm]{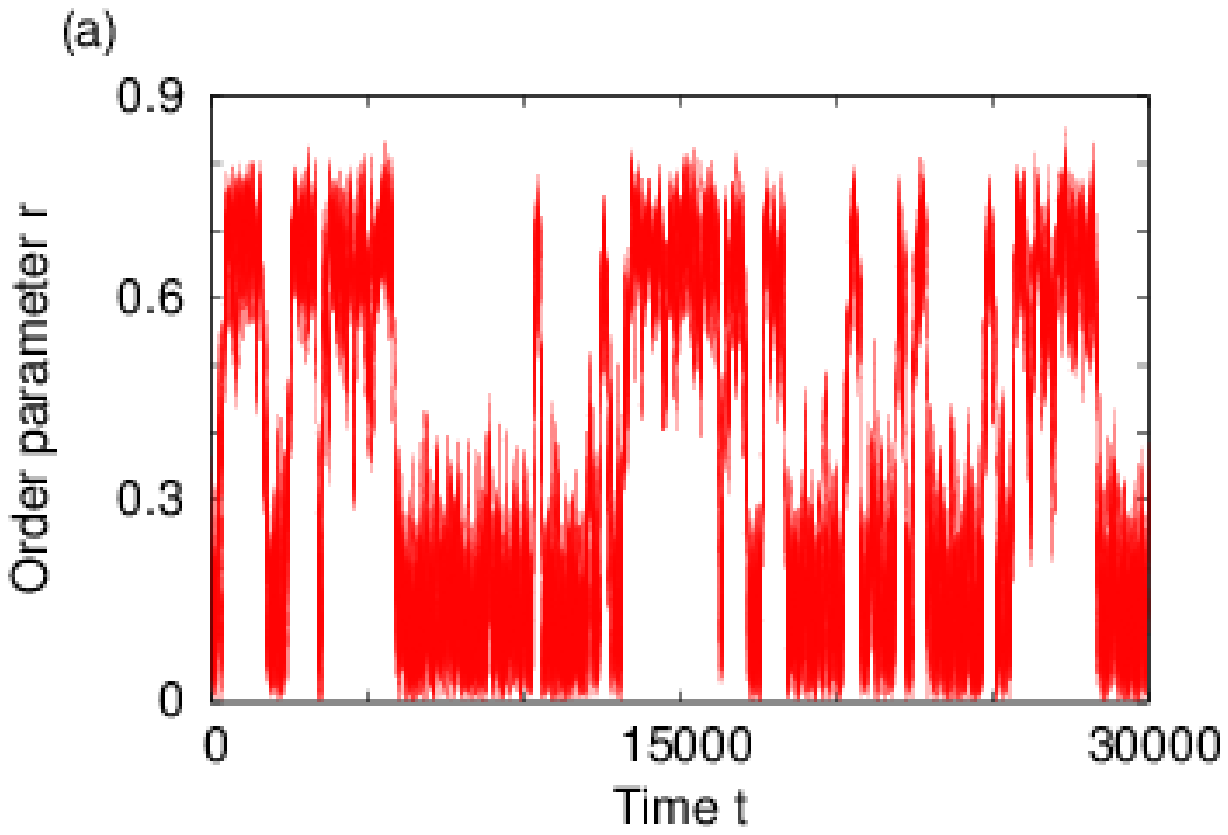} \\
\includegraphics[width=80mm]{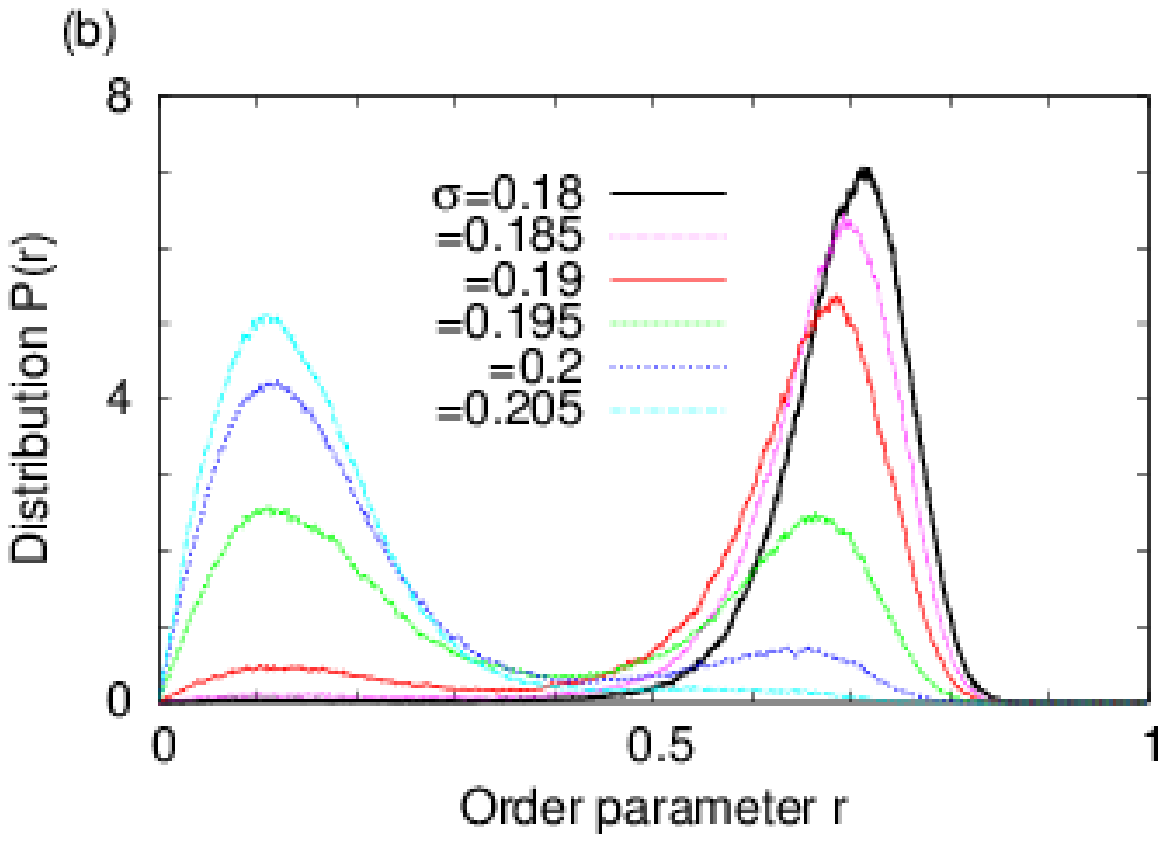}
\caption{(Color online) For $m=20,T=0.25$, and a Gaussian
$g(\omega)$ with zero mean 
and unit width, (a) shows at 
$\sigma=0.195$, the numerically estimated first-order phase transition
point, $r$ vs. time in the stationary state, while (b)
shows the distribution $P(r)$ at several $\sigma$'s 
around $0.195$. The data are obtained in $N$-body simulations with
$N=100$.}
\l{fig4ab}
\end{figure}

To confirm the first-order transition, we performed
$N$-body simulations involving integrations of Eq. 
(\ref{eom-scaled}) for a representative $g(\omega)$, i.e., a Gaussian.
Details of the simulation procedure are given in the Appendix
\ref{app-simulation}. For given $m$ and $T$ and an initial state with oscillators at $\th=0$
and angular velocities sampled from a Gaussian distribution with zero mean and
width $\propto T$, we let the system equilibrate at $\sigma=0$. We then
tune $\sigma$ adiabatically to high values and back in a cycle.  Figure
\ref{fig2} shows the behavior of the synchronization order parameter $r$ for several $m$'s at a fixed $T$
less than the BMF transition point $T_c=1/2$, illustrating sharp jumps and
hysteresis behavior expected of a first-order transition. With decrease of $m$, the jump in $r$ becomes less
sharp and the hysteresis loop area decreases, both consistent with the
transition becoming second-order-like as $m$ decreases, see Fig.
\ref{fig1}. For $m=1000$, Fig. \ref{fig2} shows $\sigma^{\rm inc}(m,T)$ and
$\sigma^{\rm coh}(m,T)$, the stability thresholds for the 
incoherent and the synchronized phase, respectively; the phase
transition point $\sigma_c(m,T)$ lies in between the two thresholds
(see
Fig. \ref{fig1}(d)). Figure \ref{fig2} shows that the thresholds decrease and approach zero with
the increase of $m$; it also suggests, together with Fig. \ref{fig3}, that $\sigma^{\rm inc}$ and $\sigma^{\rm coh}$
coincide both on the second order critical lines and as $m\to \infty$ at
a fixed $T$. 

For given $m$ and $T$ and $\sigma$ between $\sigma^{\rm
inc}(m,T)$ and
 $\sigma^{\rm coh}(m,T)$, $r$ versus time in the
stationary state shows bistability, with the system switching back and
forth between incoherent and synchronized states [Fig. \ref{fig4ab}(a)].
To have not-too-large switching times, these simulations have been performed with a relatively small number of oscillators, $N=100$, causing large fluctuations
in the order parameter $r$. Therefore, in Fig. \ref{fig4ab}(a) the
synchronized and the unsynchronized state are characterized
by values of $r$ fluctuating above and below $0.4$, respectively; however, this does allow for a clear visualization of the switches. The distribution
$P(r)$ in Fig. \ref{fig4ab}(b) is bimodal with a peak around
$r \approx 0$ or $r>0$ as $\sigma$ varies between $\sigma^{\rm inc}$ and
$\sigma^{\rm coh}$, consistent with the transition being first-order.
Indeed, a first-order transition point is characterized by two
equally likely values of the order parameter, while at a second-order
phase transition point, the order parameter has its value equal to
zero \cite{Goldenfeld:1992}.

\section{Analytical treatment}
\l{analytical}
We now turn to an analytical treatment of the first-order
transition. In the continuum limit $N \to \infty$, the dynamics Eq. (\ref{eom-scaled}) is described by the
single-oscillator distribution $f(\th,v,\omega,t)$ which
gives at time $t$ and for each $\omega$ the fraction of oscillators with phase $\th$
and angular velocity $v$. The distribution is $2\pi$-periodic in
$\theta$, and obeys the normalization $\int_0^{2\pi} \dd\th
\int_{-\infty}^{\infty} \dd v
f(\th,v,\omega,t)=1$, while evolving 
following the Kramers equation \cite{Acebron:2000}
\be
\frac{\partial f}{\partial t}=-v\frac{\partial f}{\partial
\th}+\frac{\partial}{\partial
v}\Big(\frac{v}{\sqrt{m}}-\sigma \omega-r\sin(\psi-\th)\Big)f+\frac{
T}{\sqrt{m}}\frac{\partial^2 f}{\partial v^2}, 
\l{Kramers}
\ee
where $re^{i\psi}=\int \dd\th \dd v \dd\omega
~g(\omega)e^{i\th}f(\th,v,\omega,t)$.
We now give the derivation of Eq. (\ref{Kramers}), followed by a
discussion of its stationary solution corresponding to the incoherent
phase.

\subsection{The Kramers equation for the single-oscillator
distribution: Incoherent stationary state}
Here, we start with deriving the Bogoliubov-Born-Green-Kirkwood-Yvon
(BBGKY) hierarchy equations for the dynamics Eq. (\ref{eom-scaled}) for any number of
oscillators $N$, and, from
there, by considering the limit $N \to \infty$, obtain the Kramers Eq. (\ref{Kramers}).
For simplicity, we first discuss the derivation of the BBGKY equations
for the case of a bimodal $g(\omega)$, and then generalize it to a
general $g(\omega)$.

Consider a given realization of $g(\omega)$, in which there are $N_{1}$
oscillators with frequencies $\omega_{1}$, and $N_{2}$ oscillators with
frequencies $\omega_{2}$, where $N_{1}+N_{2}=N$. 
We then define the $N$-oscillator distribution function
$f_{N}(\theta_{1},v_{1},\dots,\theta_{N_{1}},v_{N_{1}},\theta_{N_{1}+1},v_{N_{1}+1},\dots,\theta_{N},v_{N},t)$
as the probability density at time $t$ to observe the system around the
values $\{\theta_{i},v_{i}\}_{1\le i\le N}$. In the following, we use
the shorthand notations $z_{i}\equiv(\theta_{i},v_{i})$ and $\mathbf{z}=(z_{1},z_{2},\dots,z_{N})$. Note that
$f_{N}$ satisfies the normalization $\int\Big(\prod_{i=1}^{N}\dd z_{i}\Big)f_{N}(\mathbf{z},t)=1$. We assume
that 
\begin{enumerate}
\item{$f_{N}$ is symmetric with respect to permutations of dynamical variables within the group of
oscillators with the same frequency, and}
\item{$f_N$, together with the derivatives $\partial
f_{N}/\partial v_{i} ~\forall~ i$, vanish on the boundaries of the phase
space.}
\end{enumerate}

The evolution of $f_{N}$ follows the Fokker-Planck equation which may be
straightforwardly derived from the equations of motion Eq. 
(\ref{eom-scaled}): 
\begin{eqnarray}
&&\frac{\partial f_{N}}{\partial t}
=-\sum_{i=1}^{N}\Big[v_{i}\frac{\partial
f_{N}}{\partial\theta_{i}}-\frac{1}{\sqrt{m}}\frac{\partial(v_{i}f_{N})}{\partial
v_{i}}\Big] \nonumber \\
&&-\sigma\sum_{j=1}^{N}\Big(\Omega^{T}\Big)_{j}\frac{\partial
f_{N}}{\partial
v_{j}}+\frac{T}{\sqrt{m}}\sum_{i=1}^{N}\frac{\partial^{2}f_{N}}{\partial
v_{i}^{2}}\nonumber \\
&&-\frac{1}{2N}\sum_{i,j=1}^{N}\sin(\theta_{j}-\theta_{i})\Big[\frac{\partial
f_{N}}{\partial v_{i}}-\frac{\partial f_{N}}{\partial v_{j}}\Big], \label{eq:fp-eqn}
\end{eqnarray}
where we have defined the $N\times 1$ column vector $\Omega$ whose
first $N_{1}$ entries equal $\omega_{1}$ and the following $N_{2}$
entries equal $\omega_{2}$, and where the superscript $T$ denotes matrix transpose
operation: $\Omega^T\equiv\left[\omega_{1} ~\omega_{1}
\dots~\omega_{1}~\omega_{2}\dots ~\omega_{2}\right]$.

To proceed, we follow standard procedure \cite{Huang:1987}, and define the reduced
distribution function $f_{s_{1},s_{2}}$, with $s_1=0,1,2,\dots,N_{1}$
and $s_2=0,1,2,\dots,N_{2}$, as
\begin{eqnarray}
&&f_{s_{1},s_{2}}(z_{1},z_{2},\dots,z_{s_{1}},z_{N_{1}+1},\dots,z_{N_{1}+s_{2}},t)\nonumber
\\
&&=\frac{N_{1}!}{(N_{1}-s_{1})!N_{1}^{s_{1}}}\frac{N_{2}!}{(N_{2}-s_{2})!N_{2}^{s_{2}}}\nonumber
\\
&&\int \dd z_{s_{1}+1}\dots \dd z_{N_{1}}\dd z_{N_{1}+s_{2}+1}\dots
\dd z_{N}f_{N}(z,t).
\label{eq:fs-defn}
\end{eqnarray}
Note that the following normalizations
hold for the single-oscillator distribution functions: $\int 
\dd z_{1}f_{1,0}(z_{1},t)=1$, and $\int \dd z_{N_{1}+1}f_{0,1}(z_{N_{1}+1},t)=1$.

Using Eq. (\ref{eq:fp-eqn}) in Eq. (\ref{eq:fs-defn}) and
simplifying, we get the BBGKY hierarchy equations for oscillators
with frequencies $\omega_{1}$ as
\begin{eqnarray}
&&\frac{\partial f_{s,0}}{\partial t}
+\sum_{i=1}^{s}\Big[\frac{v_{i}\partial
f_{s,0}}{\partial\theta_{i}}-\frac{1}{\sqrt{m}}\frac{\partial}{\partial
v_{i}}(v_{i}f_{s,0})\Big]+\sigma\sum_{i=1}^{s}\omega_{1}\frac{\partial
f_{s,0}}{\partial
v_{i}}\nonumber \\
&&-\frac{T}{\sqrt{m}}\sum_{i=1}^{s}\frac{\partial^{2}f_{s,0}}{\partial
v_{i}^{2}}=
-\frac{1}{2N}\sum_{i,j=1}^{s}\sin(\theta_{j}-\theta_{i})\Big[\frac{\partial
f_{s,0}}{\partial v_{i}}-\frac{\partial f_{s,0}}{\partial
v_{j}}\Big]\nonumber \\
&&-\frac{N_{1}}{N}\sum_{i=1}^{s}\int \dd
z_{s+1}\sin(\theta_{s+1}-\theta_{i})\frac{\partial f_{s+1,0}}{\partial
v_{i}}\nonumber \\
&&- \frac{N_{2}}{N}\int \dd
z_{N_{1}+1}\sum_{i=1}^{s}\sin(\theta_{N_{1}+1}-\theta_{i})\frac{\partial
f_{s,1}}{\partial v_{i}},
\end{eqnarray}
and similar equations for $f_{0,s}$ for oscillators of frequencies $\omega_{2}$. The
first equations of the hierarchy are
\begin{eqnarray}
&&\frac{\partial f_{1,0}(\theta,v,t)}{\partial t} +\frac{v\partial
f_{1,0}(\theta,v,t)}{\partial\theta}-\frac{1}{\sqrt{m}}\frac{\partial}{\partial
v}(vf_{1,0}(\theta,v,t))\nonumber \\
&&+\sigma\omega_{1}\frac{\partial f_{1,0}(\theta,v,t)}{\partial v}-\frac{T}{\sqrt{m}}\frac{\partial^{2}f_{1,0}(\theta,v,t)}{\partial v^{2}}\nonumber \\
&&=-\frac{N_{1}}{N}\int \dd
 \theta'\dd v'\sin(\theta'-\theta)\frac{\partial
 f_{2,0}(\theta,v,\theta',v',t)}{\partial v}\nonumber \\
 &&-\frac{N_{2}}{N}\int \dd
 \theta'\dd v'\sin(\theta'-\theta)\frac{\partial
 f_{1,1}(\theta,v,\theta',v',t)}{\partial v},\label{eq:fp-1}
\end{eqnarray}
and
\begin{eqnarray}
&&\frac{\partial f_{0,1}(\theta,v,t)}{\partial t} +\frac{v\partial
f_{0,1}(\theta,v,t)}{\partial\theta}-\frac{1}{\sqrt{m}}\frac{\partial}{\partial
v}(vf_{0,1}(\theta,v,t))\nonumber \\
&&+\sigma\omega_{2}\frac{\partial f_{0,1}(\theta,v,t)}{\partial v}-\frac{T}{\sqrt{m}}\frac{\partial^{2}f_{0,1}(\theta,v,t)}{\partial v^{2}}\nonumber \\
&&=-\frac{N_{2}}{N}\int \dd
 \theta'\dd v'\sin(\theta'-\theta)\frac{\partial
 f_{0,2}(\theta,v,\theta',v',t)}{\partial v}\nonumber \\
 &&-\frac{N_{1}}{N}\int \dd
 \theta'\dd v'\sin(\theta'-\theta)\frac{\partial
 f_{1,1}(\theta,v,\theta',v',t)}{\partial v}.\label{eq:fp-2}
\end{eqnarray}
In the limit of large $N$, we can write
\begin{equation}
g(\omega)=
\Big[\frac{N_{1}}{N}\delta(\omega-\omega_{1})+\frac{N_{2}}{N}\delta(\omega-\omega_{2})\Big],
\end{equation}
and express Eqs. (\ref{eq:fp-1}) and (\ref{eq:fp-2}) in terms of
$g(\omega)$.

In order to generalize Eqs. (\ref{eq:fp-1}) and (\ref{eq:fp-2}) to the
case of a continuous $g(\omega)$, we denote for this case the
single-oscillator distribution function as $f(\theta,v;\omega,t)$. The first
equation of the hierarchy is then
\begin{eqnarray}
&&\frac{\partial f(\theta,v,\omega,t)}{\partial t} +\frac{v\partial
f(\theta,v,\omega,t)}{\partial\theta}-\frac{1}{\sqrt{m}}\frac{\partial}{\partial
v}(vf(\theta,v,\omega,t))\nonumber \\
&&+\sigma\omega\frac{\partial f(\theta,v,\omega,t)}{\partial
v}-\frac{T}{\sqrt{m}}\frac{\partial^{2}f(\theta,v,\omega,t)}{\partial v^{2}}\nonumber \\
&&=  -\int \dd \omega'g(\omega')\int \dd
\theta'\dd v'\sin(\theta'-\theta)\frac{\partial
f(\theta,v,\theta',v',\omega,\omega',t)}{\partial v}.\nonumber \\\label{eq:final-eqn}
\end{eqnarray}

In the continuum limit $N\to\infty$, we may neglect two-oscillator
correlations and approximate $f(\theta,v,\theta',v',\omega,\omega',t)$
as
\begin{eqnarray}
&&f(\theta,v,\theta',v',\omega,\omega',t)
=f(\theta,v,\omega,t)f(\theta',v',\omega',t)\nonumber \\
&&+\mathrm{~corrections
~subdominant ~in}~N,
\l{twop}
\end{eqnarray}
so that Eq. (\ref{eq:final-eqn}) reduces to the Kramers Eq. (\ref{Kramers}).

The stationary solutions of Eq. (\ref{Kramers}) are obtained
by setting the left hand side to zero. For $\sigma=0$, the stationary
solution is 
\be
f_{\rm st}(\th,v)\propto \exp[-(v^2/2-r_{\rm
st}\cos \th)/T],
\ee
that corresponds to canonical equilibrium, with $r_{\rm st}$ determined
self-consistently  \cite{Chavanis:2013}. 
For $\sigma \ne 0$, the incoherent stationary state is
\cite{Acebron:2000} 
\be
f^{\rm inc}_{\rm
st}(\th,v,\omega)=\fr{1}{(2\pi)^{3/2}\sqrt{T}}\exp[-(v-\sigma \omega
\sqrt{m})^2/(2T)].
\l{inc-st}
\ee
The existence of the synchronized stationary state is borne out by our
simulation results discussed above (see Figs.
\ref{fig2},\ref{fig3}, and \ref{fig4ab}), although its analytical form is not
known.

\subsection{Linear stability analysis of
the incoherent state}
Let us now discuss the linear stability analysis of
the incoherent state Eq. (\ref{inc-st}), pursued in Ref.
\cite{Acebron:2000} by linearizing Eq. (\ref{Kramers}) about the state by expanding
$f$ as 
\be
f(\th,v,\omega,t)=f^{\rm inc}_{\rm st}(\th,v,\omega)+e^{\lambda
t}\delta f(\th,v,\omega),
\ee
with $\delta f \ll 1$. The solution of
the linearized equation yields that $\lambda$
satisfies \cite{Acebron:2000} 
\be
\fr{2T}{e^{mT}}=\sum_{p=0}^\infty
\frac{(-m T)^p(1+\frac{p}{mT})}{p!}\int\limits_{-\infty}^{\infty}
\frac{g(\omega)\dd\omega}{
1+\frac{p}{mT}+i\frac{\sigma\omega}{T}+\frac{\lambda}{T\sqrt{m}}}.
\l{stability-eqn}
\ee
The above equation contains valuable information about the
range of values of the parameters $m,T,\sigma$ for which the incoherent
state is stable, and consequently, about the
transition from the incoherent to synchronized phase. This warrants
a detailed analysis of Eq. (\ref{stability-eqn}) for a general unimodal 
$g(\omega)$. The
analysis for Lorentzian $g(\omega)$ in Ref.
\cite{Acebron:2000} left untouched the crucial issue of the
synchronization transition.

We rewrite Eq. (\ref{stability-eqn}) as
\begin{eqnarray}
\label{eigeneq}
&&F(\lambda;m,T,\sigma) \equiv \frac{e^{mT}}{2T}\sum_{p=0}^\infty \frac{\left(-mT\right)^p
\left(1+\frac{p}{mT}\right)}{p!}\nonumber \\
&&\times \int \frac{g(\omega)\dd \omega}{1+\frac{p}{mT}+\frac{\lambda}{T\sqrt{m}} +
i\frac{\sigma \omega}{T}} - 1 = 0,
\end{eqnarray}
where $g(\omega)$ is unimodal. 
The incoherent state is unstable if there is a $\lambda$ with a positive
real part that satisfies the above eigenvalue equation. We will now prove that, depending on the values of the parameters
appearing in the above equation, there can be at most one such $\lambda$
that can be only real. In addition, for the case of a Gaussian $g(\omega)$ explicitly
used in simulations reported in this paper,
we obtain the general shape of the surface in the $(m,T,\sigma)$ space
that defines the instability region of the incoherent state.
\begin{figure}[here!]
\centering
\includegraphics[width=60mm,angle=270]{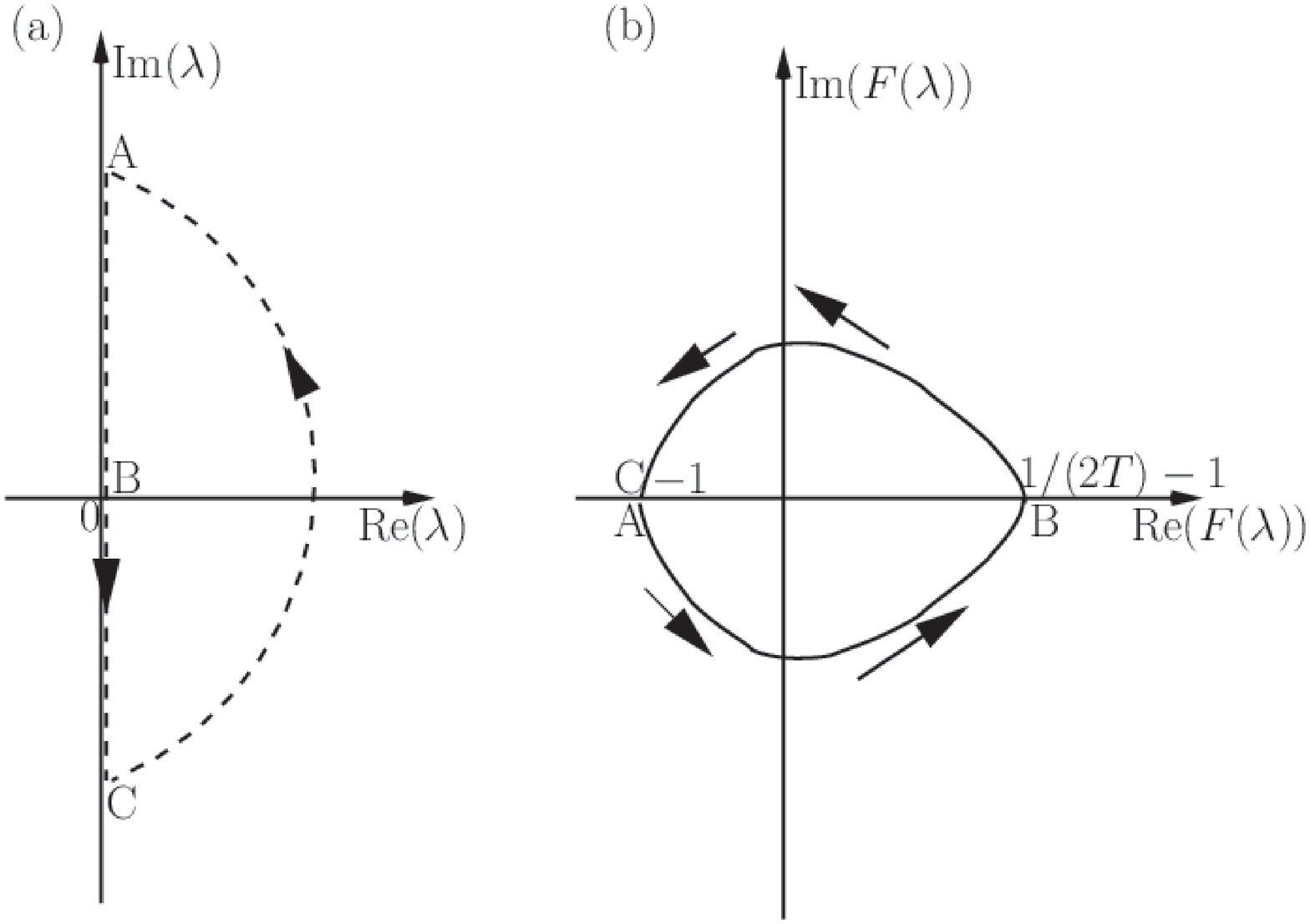}
\caption{The loop in the complex $F$-plane, (b), corresponding to the
loop in the complex $\lambda$-plane, (a), as determined by the function
$F(\lambda)$ in Eq. (\ref{eqfond}).}
\label{sm-fig1}
\end{figure}

Considering $m$ and $T$ strictly positive, we multiply for convenience
the numerator and denominator of Eq. (\ref{eigeneq}) by $mT$ to obtain
\begin{eqnarray}
\label{eqfond}
&&F(\lambda;m,T,\sigma) = \frac{e^{mT}}{2T}\sum_{p=0}^\infty
\frac{\left(-mT\right)^p \left(p+mT\right)}{p!}\nonumber \\
&&\times \frac{g(\omega)\dd\omega}{mT+p+\sqrt{m}\lambda + i\sigma m \omega} - 1 = 0.
\end{eqnarray}
Let us first look for pure imaginary solutions of this equation.
Separating into real and imaginary parts, we have
\begin{eqnarray}\label{iml_real}
&&{\rm Re} \left[F(i\mu;m,T,\sigma)\right]=
\frac{e^{mT}}{2T}\sum_{p=0}^\infty
\frac{\left(-mT\right)^p}{p!}\nonumber \\
&&\times \int \dd \omega \, g(\omega)
\frac{\left(p+mT\right)^2}{\left(p+mT\right)^2+\left(m\sigma \omega
+\sqrt{m}\mu\right)^2} - 1 = 0, \nonumber \\ \\
\label{iml_imag}
&&{\rm Im} \left[F(i\mu;m,T,\sigma)\right]= -\frac{e^{mT}}{2T}
\sum_{p=0}^\infty \frac{\left(-mT\right)^p}{p!}\nonumber \\
&&\times\int \dd \omega \, g(\omega) \frac{\left(p+mT\right)\left(m\sigma \omega + \sqrt{m}\mu \right)}
{\left(p+mT\right)^2+\left(m\sigma \omega + \sqrt{m}\mu\right)^2}= 0.
\end{eqnarray}
In the second equation above, we make the change of variables $m\sigma
\omega + \sqrt{m}\mu = m\sigma x$, and exploit the parity in $x$ of the sum,
to obtain
\begin{eqnarray}\label{iml_imag_2}
&&{\rm Im} \left[F(i\mu;m,T,\sigma)\right]=\nonumber \\
&&-m\sigma\int_0^\infty \dd
x\Big\{ \left[g\left(x-\frac{\mu}{\sqrt{m}\sigma}\right)
-g\left(-x-\frac{\mu}{\sqrt{m}\sigma}\right)\right]\nonumber \\
&&\times x \sum_{p=0}^\infty \frac{\left(-mT\right)^p}{p!}
\frac{p+mT}{\left(p+mT\right)^2+m^2\sigma^2 x^2} \Big\} = 0.
\end{eqnarray}
It can be shown that the sum on the
right-hand side is positive definite for any finite $\sigma$. Furthermore,
for our class of distribution functions, one may see that the term in
square brackets is positive (respectively, negative) definite for $\mu >0$
(respectively, for $\mu <0$). As a consequence, the last equation is never
satisfied for $\mu \ne 0$ and finite, and therefore, the eigenvalue equation does not admit
pure imaginary solutions [the proof holds also for the particular case
$g(\omega) = \delta(\omega)$, as may be checked].
We also conclude that there can be at most one solution with positive
real part. In fact, if in the complex $\lambda$-plane, we
perform the loop depicted in Fig. \ref{sm-fig1}(a) (where it is
meant that
the points $A$ and $C$ represent ${\rm Im}\lambda \to \pm \infty$, respectively, and the radius of the arc
extends to $\infty$), then, in the
complex-$F(\lambda)$ plane, we obtain, due to the sign properties
of ${\rm Im} \left[F(i\mu;m,T,\sigma)\right]$ just described, the loop
qualitatively represented in Fig. \ref{sm-fig1}(b). The point $F=-1$ in
Fig. \ref{sm-fig1}(b) is obtained for $\lambda$, in Fig. \ref{sm-fig1}(a), for values at points $A$ and $C$
and in the whole of the arc extending to infinity. The position of the point $B$ in the complex-$F$ plane is
determined by the value of $F(0)$, which is given by
\begin{eqnarray}\label{rel_real}
&&F(0;m,T,\sigma) = \frac{e^{mT}}{2T}\sum_{p=0}^\infty
\frac{\left(-mT\right)^p}{p!}\nonumber \\
&&\int \dd \omega \, g(\omega)
\frac{\left(p+mT\right)^2}{\left(p+mT\right)^2+\left(m\sigma \omega \right)^2} - 1.
\end{eqnarray}
From  the well-known theorem of complex analysis on the number of roots
of a function in a given domain of the complex plane \cite{Smirnov:1964}, we therefore obtain that
for $F(0;m,T,\sigma)>0$, there is one and only one solution of the
eigenvalue equation with positive real part; on the other hand, for
$F(0;m,T,\sigma)<0$, there is no such solution. When the single solution with
positive real part exists, it is necessarily real, since a complex solution
would imply the existence of its complex conjugate. The value of
$F(0;m,T,\sigma)$ is readily seen to be equal to $1/(2T)-1$ for
$\sigma = 0$. For positive $\sigma$, 
the value will depend on the particular form of the distribution function $g(\omega)$. However, it is possible to prove that the value is
always smaller than $1/(2T)-1$; this is consistent with the physically reasonable fact that if the incoherent state is stable for
$\sigma =0$, which happens for $T>1/2$, it is all the more stable for $\sigma > 0$.

The surface delimiting the region of instability in the $(m,T,\sigma)$
phase space is implicitly defined by Eq. (\ref{rel_real}) [i.e.
$F(0;m,T,\sigma)=0$], which, in principle, can be solved to obtain
the threshold value of $\sigma$ (denoted by $\sigma^{\rm inc}$) as a function of
$(m,T)$: $\sigma^{\rm inc}=\sigma^{\rm inc}(m,T)$. On physical grounds,
we expect that the latter is a single valued function, and that for any
given value of $m$, it is a decreasing function of $T$ for $0\le T \le
1/2$, reaching $0$ for $T=1/2$. We are able to prove
analytically these facts for the class of unimodal distribution
functions
$g(\omega)$ considered in this work that includes
the Gaussian case. However, we can prove in general for any $g(\omega)$ that $\sigma^{\rm inc}(m,T)$ tends to
$0$ for $m\to \infty$. This is done using the integral representation
\bea\label{integr_repr_2}
&&\sum_{p=0}^\infty \frac{\left(-mT\right)^p}{p!} \frac{\left(p+mT\right)^2}{\left(p+a\right)^2+\left(m\sigma \omega \right)^2}=
e^{-mT}\nonumber \\
&&- \left(m\sigma \omega \right) \int_0^\infty \dd t \, \exp \left[ -mT \left( t + e^{-t}\right)\right]\sin \left(m\sigma \omega t \right).
\eea
For $\sigma>0$ and $m\to \infty$, one may see that the term within the
integral in the last equation tends to $e^{-mT}$. We thus obtain by
examining Eq. (\ref{rel_real}) that $F(0;m\to \infty,T>0,\sigma>0)=-1$.
Combined with the fact that $F(0;m,T,0)=1/(2T)-1$, this shows that
$\sigma^{\rm inc}(m\to \infty,0\le T \le 1/2)=0$.

Let us now turn to the Gaussian case, $g(\omega)= \frac{1}{\sqrt{2\pi}} \exp \left[-\frac{\omega^2}{2}\right]$. Denoting with a
subscript $g$ in this case, and using Eq. (\ref{integr_repr_2}), we have
\begin{eqnarray}\label{rel_real_b_gauss}
&&F_g(0;m,T,\sigma) = \frac{1}{2T}- 1-\frac{e^{mT}}{2T\sqrt{2\pi}}\int \dd
\omega \, e^{-\frac{\omega^2}{2}} \left(m\sigma \omega \right)\nonumber
\\
&&\times \int_0^\infty \dd t \, \exp \left[ -mT \left( t + e^{-t}\right)\right]
\sin \left(m\sigma \omega t \right).
\end{eqnarray}
The integral in $\omega$ can be easily performed. Making the change of
variable $m\sigma t=y$, we arrive at the following equation:
\begin{eqnarray}\label{rel_real_b_gauss_3}
&&F_g(0;m,T,\sigma) = \frac{1}{2T}-1\nonumber \\
&&- \frac{1}{2T}\int_0^\infty \dd y \,
y e^{-\frac{y^2}{2}}\exp \left[ mT \left( 1 - \frac{y}{m\sigma} -
e^{-\frac{y}{m\sigma}}\right)\right].
\end{eqnarray}
The equation $F_g(0;m,T,\sigma)=0$ defines implicitly the function
$\sigma^{\rm inc}(m,T)$. We can show that this is a single-valued
function with the properties $\frac{\partial \sigma^{\rm inc}}{\partial m}<0$ and $\frac{\partial \sigma^{\rm inc}}{\partial T}<0$.
We show this by explicitly computing the partial derivatives of $F_g(0;m,T,\sigma)$ with respect to
$m$ and $\sigma$, and by evaluating the behavior with respect to changes
in $T$ by adopting a suitable strategy.
We begin by computing the derivative with respect to $\sigma$. From Eq.
(\ref{rel_real_b_gauss_3}), we readily obtain
\begin{eqnarray}\label{rel_real_b_gauss_dersigma}
&&\frac{\partial}{\partial \sigma}F_g(0;m,T,\sigma)= - \frac{1}{2\sigma^2}\int_0^\infty \dd y \,
y^2 e^{-\frac{y^2}{2}} \left( 1- e^{-\frac{y}{m\sigma}} \right)\nonumber
\\
&&\times\exp \left[ mT \left( 1 - \frac{y}{m\sigma}
- e^{-\frac{y}{m\sigma}}\right)\right],
\end{eqnarray}
which is clearly negative. Second, the derivative with respect to $m$ gives
\begin{eqnarray}\label{rel_real_b_gauss_dermass}
&&\frac{\partial}{\partial m}F_g(0;m,T,\sigma) = - \frac{1}{2}\int_0^\infty \dd y \,
y e^{-\frac{y^2}{2}}\nonumber \\
&&\times\left( 1-
e^{-\frac{y}{m\sigma}} - \frac{y}{m\sigma}e^{-\frac{y}{m\sigma}}
\right) \exp \left[ mT \left( 1 - \frac{y}{m\sigma} -
e^{-\frac{y}{m\sigma}}\right)\right]. \nonumber \\
\end{eqnarray}
This derivative is negative, since $1-e^{-x} -xe^{-x}$ is positive for
$x>0$. From the implicit function theorems, we then
derive that $\frac{\partial \sigma^{\rm inc}}{\partial m}<0$. The study of the behavior with respect to a change in $T$ is a bit
more complicated. Since we are considering $T>0$, we multiply Eq.
(\ref{rel_real_b_gauss_3}) by $2T$ to obtain
\begin{eqnarray}\label{rel_real_b_gauss_4}
&&2T F_g(0;m,T,\sigma) = 1 - 2T \nonumber \\
&&- \int_0^\infty \dd y \,
y e^{-\frac{y^2}{2}}\exp \left[ mT \left( 1 - \frac{y}{m\sigma} - e^{-\frac{y}{m\sigma}}\right)\right].
\end{eqnarray}
Let us consider the integral on the right-hand side
\begin{equation}\label{rel_real_b_gauss_4_b}
\int_0^\infty \dd y \,
y e^{-\frac{y^2}{2}} \exp \left[ mT \left( 1 - \frac{y}{m\sigma} - e^{-\frac{y}{m\sigma}}\right)\right].
\end{equation}
Since $1-x-e^{-x}$ is negative for $x>0$, we conclude that the $T$ derivative of this expression
is negative, while its second $T$ derivative is positive. Then the right-hand side of Eq.
(\ref{rel_real_b_gauss_4}) can be zero, for $T>0$, for at most one value of $T$. Furthermore, since for
fixed $y$ and $m$ the value of $y/(m\sigma)$ decreases if $\sigma$ increases, the $T$ value for which
$F_g(0;m,T,\sigma)=0$ decreases for increasing $\sigma$ at fixed $m$.
This concludes the proof. Furthermore, for what we have seen before,
$\sigma^{\rm inc}(m,1/2)=0$ and
$\lim_{m\to \infty}\sigma^{\rm inc}(m,T)=0$ for $0\le T \le 1/2$.

From the above analysis, it should be clear that the proof is not
restricted to the Gaussian case, but would work exactly in the same way for any $g(\omega)$ such that
\begin{equation}\label{intsin_gen}
\beta \int \dd x \, g(x) x \sin (\beta x),
\end{equation}
is positive for any $\beta$. However, on physical grounds, we are led to assume that the same conclusions
hold for any unimodal $g(\omega)$. 

On the basis of our analysis, it follows that at the point of neutral
stability, one has $\lambda=0$,
which when substituted in Eq. (\ref{stability-eqn}) gives $\sigma^{\rm inc}(m,T)$ to be satisfying 
\be
\frac{2T}{e^{mT}}=\sum_{p=0}^\infty
\frac{(-mT)^p(1+\frac{p}{mT})^2}{p!}\int\limits_{-\infty}^{\infty}
\frac{g(\omega)\dd\omega}{(1+\frac{p}{mT})^2+\frac{(\sigma^{\rm
inc})^2\omega^2}{T^2}}.
\l{sigma_inc}
\ee
In the $(m,T,\sigma)$ space, Eq. (\ref{sigma_inc}) defines the stability surface $\sigma^{\rm
inc}(m,T)$. There will similarly be the stability surface $\sigma^{\rm
coh}(m,T)$ (see Fig. \ref{fig1}(d) which shows the two surfaces as
obtained in $N$-body simulations for $N=500$ for a Gaussian $g(\omega)$). The two surfaces coincide on the critical lines on the
$(T,\sigma)$ and $(m,T)$ planes where the transition becomes continuous;
outside these planes, the surfaces enclose the first-order transition surface
$\sigma_c(m,T)$ i.e., $\sigma^{\rm coh}(m,T) > \sigma_c(m,T) >\sigma^{\rm
inc}(m,T)$. We now show
by taking limits that the surface $\sigma^{\rm
inc}(m,T)$ meets the critical lines on the $(T,\sigma)$ and
$(m,T)$ planes, and also obtain its intersection with the $(m,\sigma)$-plane.
On considering $m \to 0$ at a fixed $T$, only the $p=0$ term in the
sum in Eq. (\ref{sigma_inc}) contributes, giving 
\be
\lim_{m \to 0, T \,{\rm fixed}}\sigma^{\rm inc}(m,T)=\sigma_c(m=0,T),
\l{saka-limit}
\ee
with the implicit expression of $\sigma_c(m=0,T)$ given by Eq.
(\ref{saka-result}). Similarly, one finds that 
\be
\lim_{T \to T_c^-, m \,{\rm
fixed}}\sigma^{\rm inc}(m,T)=0,
\l{bmf-limit}
\ee
that is, on the $(m,T)$ plane, the transition line
is given by $T_c=1/2$.
When $T \to 0$ at a fixed $m$, we get 
\be
\sigma^{\rm inc}_{\rm noiseless}(m)\equiv \lim_{T \to 0, m \,{\rm fixed}}\sigma^{\rm
inc}(m,T),
\l{tanaka-limit}
\ee
with 
\bea
&&1=\fr{\pi g(0)}{2\sigma^{\rm inc}_{\rm noiseless}}-\fr{m}{2}\int_{-\infty}^{\infty}\fr{g(\omega)\dd\omega}{1+m^2(\sigma^{\rm
inc}_{\rm noiseless})^2\omega^2}.
\eea
\begin{figure}[here]
\centering
\includegraphics[width=80mm]{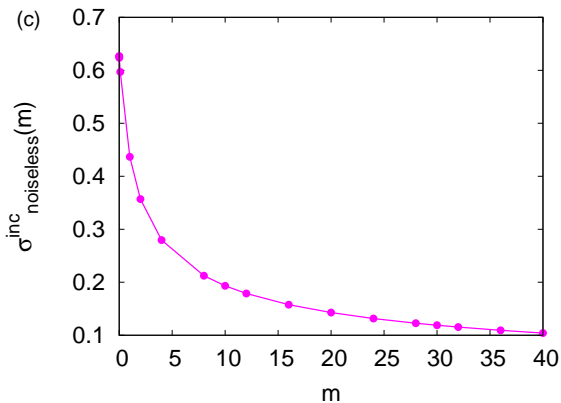}
\caption{(Color online) The figure shows the limit Eq. (\ref{tanaka-limit})
for the case of a Gaussian $g(\omega)$ with zero mean and unit width.} 
\l{sigma-limits}
\end{figure}
For the case of a Gaussian $g(\omega)$, the limits (\ref{saka-limit}) and (\ref{bmf-limit}) are
shown in Figs. \ref{fig1}(b) and \ref{fig1}(c), while the limit Eq. (\ref{tanaka-limit}) is shown in Fig. \ref{sigma-limits}.

\section{Comparison with numerics}
\l{numerical}
For a Gaussian $g(\omega)$, Eq. (\ref{sigma_inc}) gives
\be
1=\frac{e^{mT}\sqrt{\pi}}{2\sqrt{2}\sigma^{\rm inc}} \sum_{p=0}^\infty
\frac{(-mT)^p(1+\frac{p}{mT})}{p!e^{-\frac{T^2(1+p/mT)^2}{2(\sigma^{\rm
inc})^2}}}{\rm
Erfc}\Big[\frac{T(1+\frac{p}{mT})}{\sigma^{\rm inc}\sqrt{2}}\Big].
\l{Gaussian-eqn}
\ee
Choosing $m=20$ and $T=0.25$, the above equation gives $\sigma^{\rm inc}(m,T) \approx
0.10076$. Then, preparing the system in the
incoherent stationary state at a given $\sigma$, our
theoretical analysis predicts that $r$, 
in the dynamically unstable regime of the incoherent state
(i.e., with $\sigma < \sigma^{\rm
inc}$), relaxes at long times to its steady-state
value corresponding to the synchronized phase. For $\sigma >
\sigma^{\rm inc}(m,T)$, when the incoherent initial state is
linearly stable, $r$ is zero for all times.
We now compare the above continuum-limit predictions with $N$-body
simulations. We monitor the evolution of $r$ in time while
starting from the incoherent stationary state. To discuss the results,
we employ the standard picture of phase transitions occurring
dynamically as the dissipative relaxation of the order parameter towards
the minimum of a phenomenological Landau free energy \cite{Binder:1987}.
For a first-order phase transition, we draw in Fig. \ref{fig5} the
corresponding schematic free energy $F(r)$ versus $r$ for fixed $m$ and $T$
at different $\sigma$'s \cite{footnote}. The picture helps to
explain, e.g., the flips in $r$ in Fig. \ref{fig4ab}, which correspond to dynamics at $\sigma$ close to $\sigma_c$,
when the system switches back and forth between the two almost stable
synchronized and incoherent states. 
\begin{figure}[here!]
\centering \includegraphics[width=90mm]{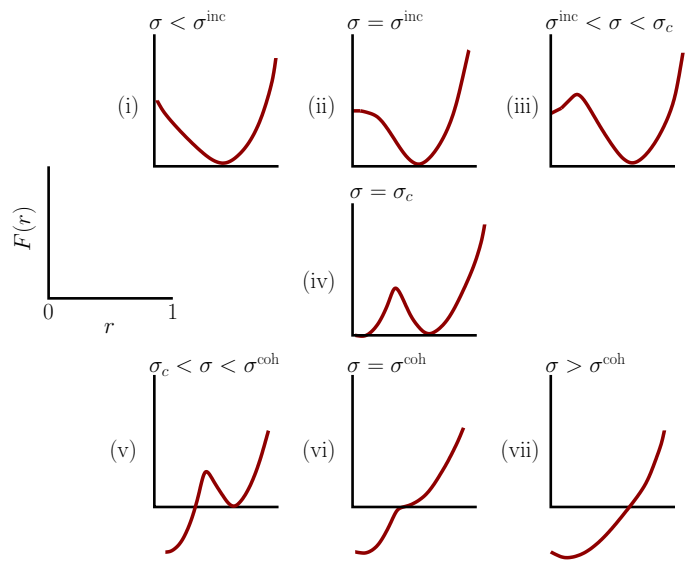}\\
\caption{(Color online) Schematic Landau free energy $F(r)$ versus $r$ for first-order
transitions at fixed $m$ and $T$ while varying $\sigma$.
Panels (i) and (vii) correspond to the synchronized and incoherent 
phase being at the global minimum. In panel (iii) (respectively, (v)), the
synchronized (respectively, incoherent) phase is at the global
minimum, while the incoherent (respectively, synchronized) phase is at a
local minimum, hence, metastable. Panel (iv) corresponds to the first-order transition point,
with the two phases coexisting at two minima of equal heights.}
\l{fig5}
\end{figure}

\begin{figure}[here!]
\centering
\includegraphics[width=60mm]{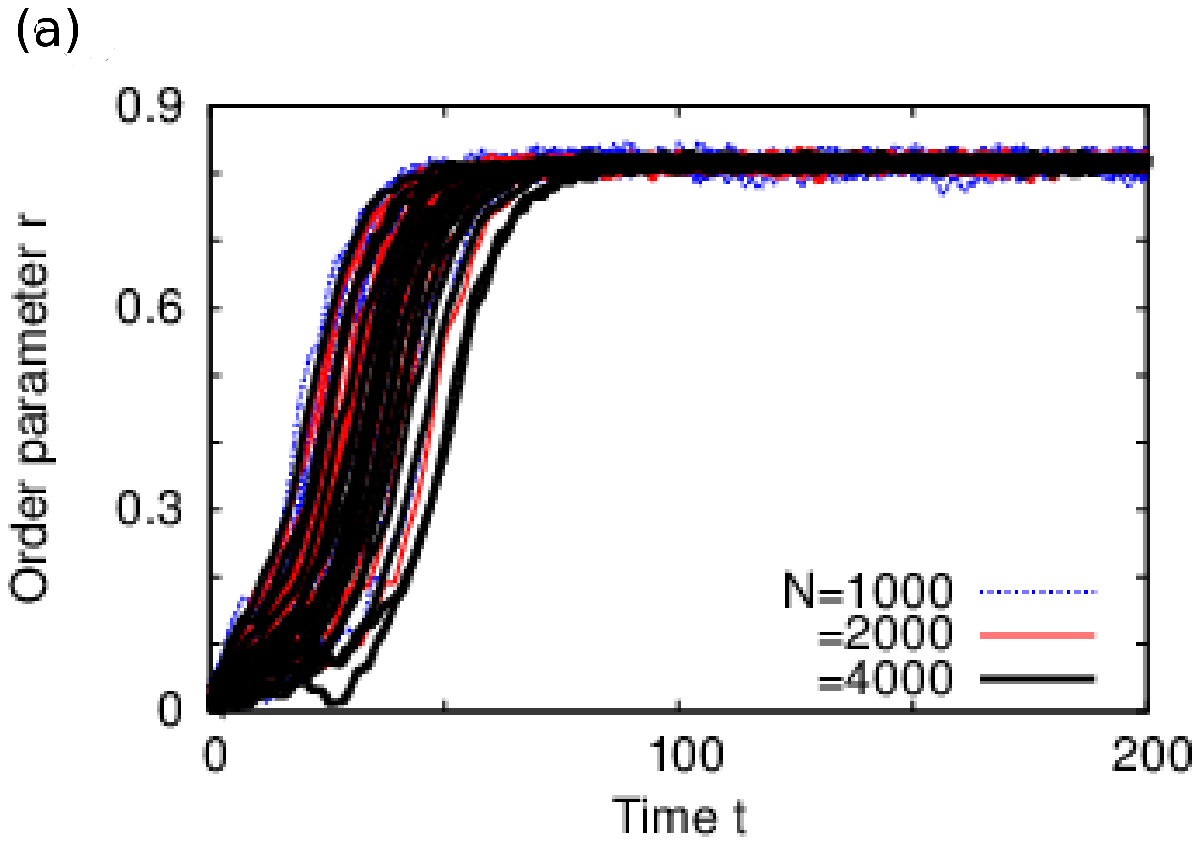}
\includegraphics[width=60mm]{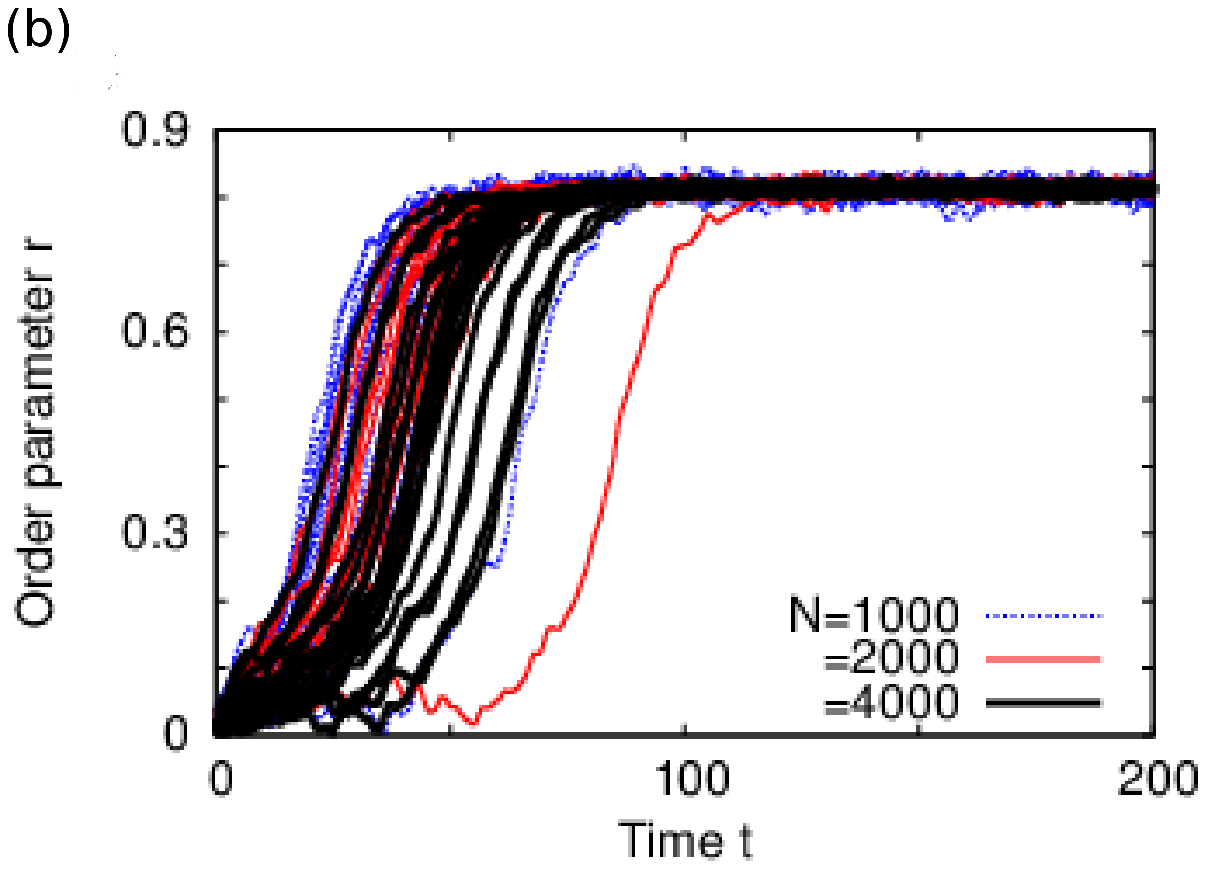}
\includegraphics[width=60mm]{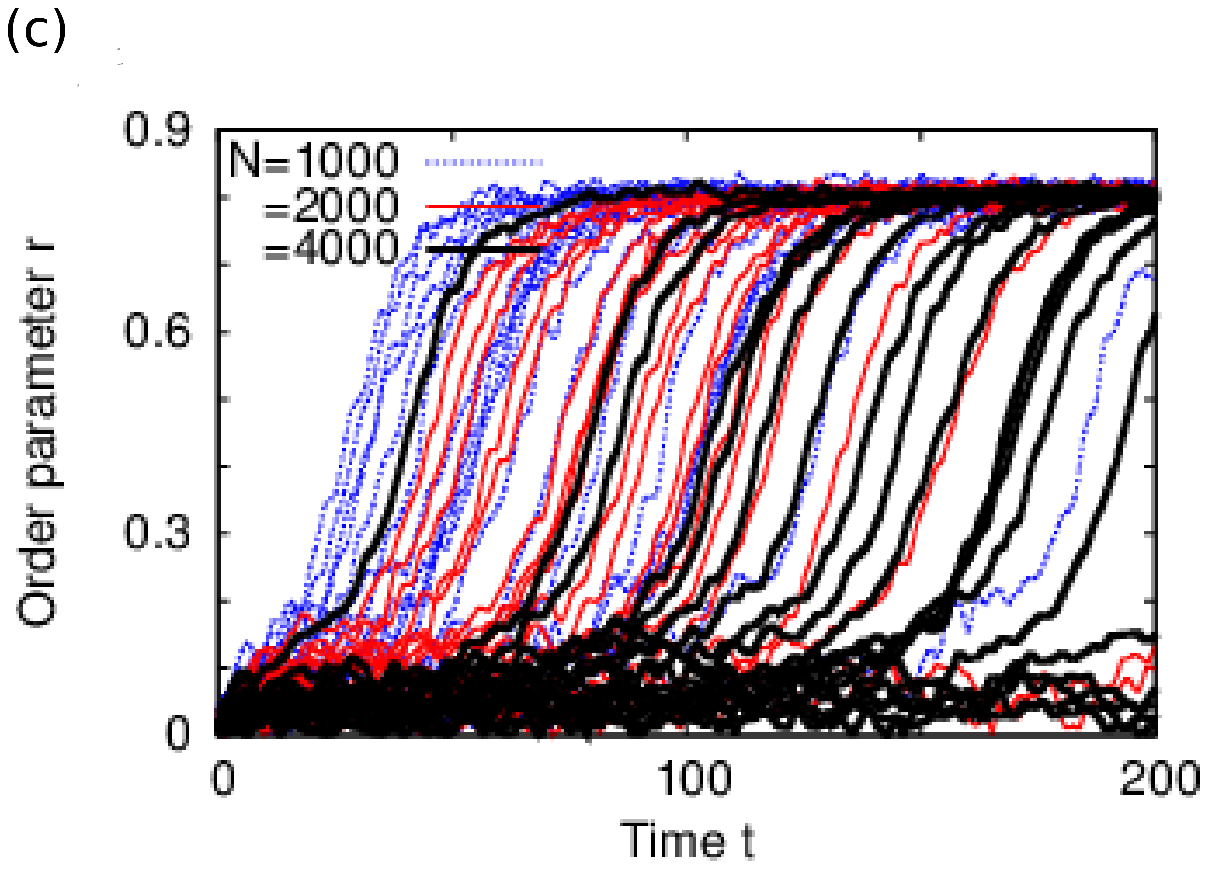}
\includegraphics[width=60mm]{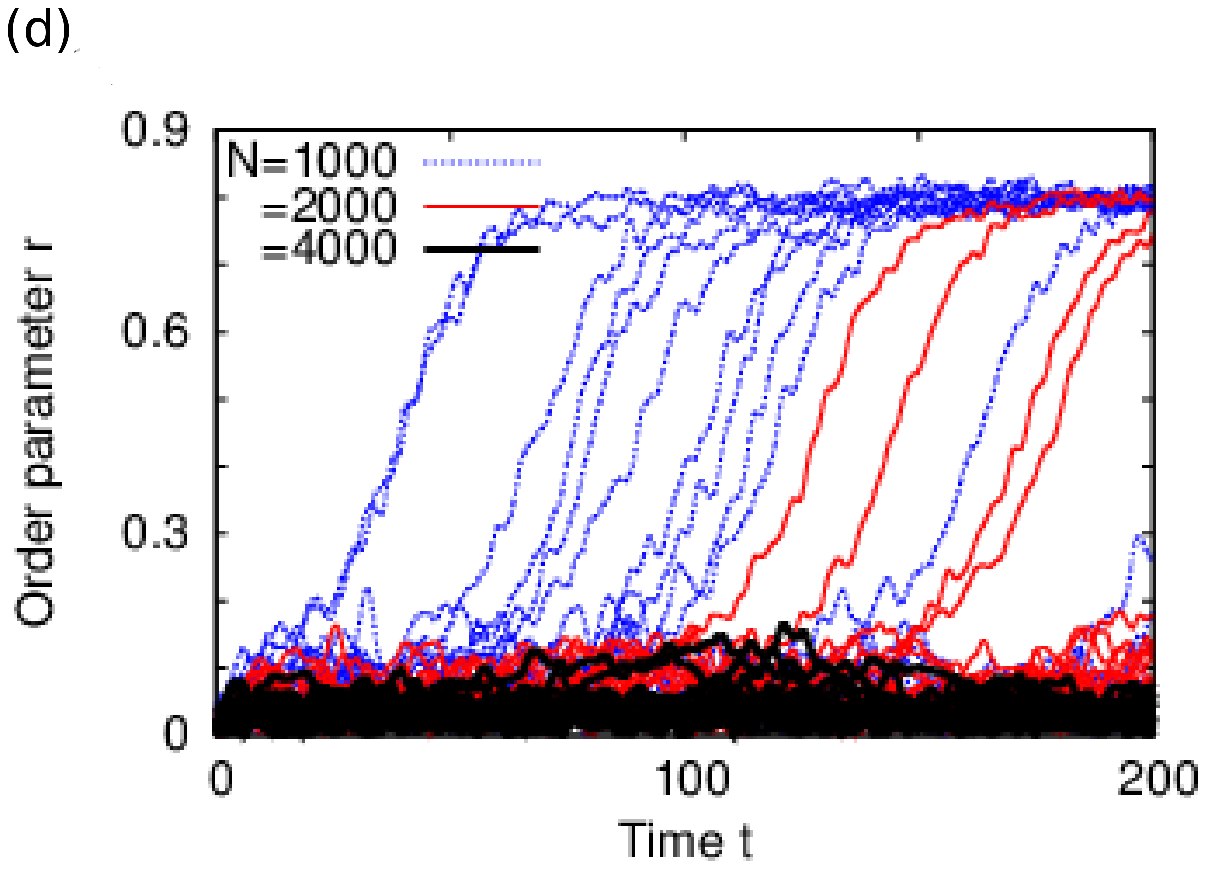}
\caption{(Color online) 
Panels
(a)-(d) show $r$ vs. time at $m=20, T=0.25$ for four values of
$\sigma$, two below ((a): $\sigma=0.09$, (b): $\sigma=0.095$), and two
above ((c): $\sigma=0.11$, (d): $\sigma=0.12$) the theoretical
threshold $\sigma^{\rm inc}(m,T)\approx 0.10076$. The data are obtained in
$N$-body simulations for a
Gaussian $g(\omega)$ with zero mean and unit width.}
\l{fig6}
\end{figure}

Let us investigate the
dynamics for $\sigma$ around $\sigma^{\rm inc}(m,T)$. Figures
\ref{fig6}(a)-(d) show simulation results for $r$ versus time for four
values of $\sigma$, two below and two above $\sigma^{\rm inc}(m,T)$. In each
case, we display the dependence for $20$ realizations of
the initial incoherent state for three values of $N$. 
Figure \ref{fig6}(a) for $\sigma < \sigma^{\rm inc}(m,T)$ illustrates that the system
while starting from the unstable incoherent state settles down in
time into the globally stable synchronized state; this corresponds to
dynamics in the landscape in Fig. \ref{fig5}(i). The relaxation of $r$ from the initial to final
synchronized state value occurs exponentially fast in time as $e^{\lambda
t}$; the growth rate $\lambda$ is obtained from Eq.
(\ref{stability-eqn}) after substituting a Gaussian
distribution for $g(\omega)$. In Fig. \ref{fig7}, we
demonstrate a match of $\lambda$ in theory and simulations. 

In Fig. \ref{fig6}(b), when $\sigma$ is larger than in Fig. \ref{fig6}(a), yet below
$\sigma^{\rm inc}(m,T)$, the system settles at long times into the synchronized
state for all realizations. Yet, some of them, at short times, tend to stay in the
initial incoherent state due to finite-$N$ effects not captured by our
continuum limit theory; see Eq. (\ref{twop}). For $\sigma > \sigma^{\rm inc}(m,T)$, but
$\sigma < \sigma_c(m,T)$, we expect on the basis of the
landscape sketched in Fig. \ref{fig5}(iii) that the
system settles at long times into the globally stable synchronized state, while for finite
times, remains trapped in the metastable incoherent state. Indeed,
Fig. \ref{fig6}(c) shows that most realizations relax to synchronized states. However,
as $N$ increases, the number of realizations staying close to the
initial incoherent state for a finite time increases. We found that the
fraction $\eta$ of realizations relaxing to synchronized
state within a fixed time decreases exponentially fast in $N$ for large
$N$; see Fig.
\ref{fig8}. This observation 
implies that for the fixed time of observation,
there exists a larger $N$ than the ones in Fig. \ref{fig6}(c) for which all realizations
remain close to the incoherent state; it then follows that in the continuum limit, all
realizations stay close to the incoherent state.

To explain the above mentioned behavior of $\eta$ with $N$, let us first consider the noisy dynamics of a single particle on a
potential landscape, when the typical time to get 
out of a metastable state is given in the weak-noise
limit by the Kramers time, i.e., an exponential in the ratio of the potential energy barrier to come
out of the metastable state to the strength of the noise
\cite{Kramers:1940}. For the dynamics of the
order parameter on a free-energy landscape for mean-field systems, the
escape time out of a metastable state obeys Kramers formula with the value of
the free-energy barrier replacing the potential energy barrier, and with an extra 
factor of $N$ multiplying the barrier height \cite{Griffiths:1966}; this
explains Fig. \ref{fig6}(c) and the behavior of $\eta$.

Figure \ref{fig6}(d), for $\sigma$ larger than $\sigma^{\rm inc}(m,T)$
than in Fig. \ref{fig6}(c), shows that with respect to (c), more realizations stay close to the initial incoherent
state for longer times, due to a larger barrier between the incoherent
and synchronized state. 
On the basis of the above discussions, we conclude that our theoretical
predictions are borne out by our simulation results. In
particular, the simulation results for $N=500$ suggest that the stability threshold of the
incoherent state lies in between $\sigma=0.095$ and
$\sigma=0.11$, a range that includes its theoretical continuum-limit value ($\approx
0.10076$). 

\begin{figure}[here]
\centering
\includegraphics[width=80mm]{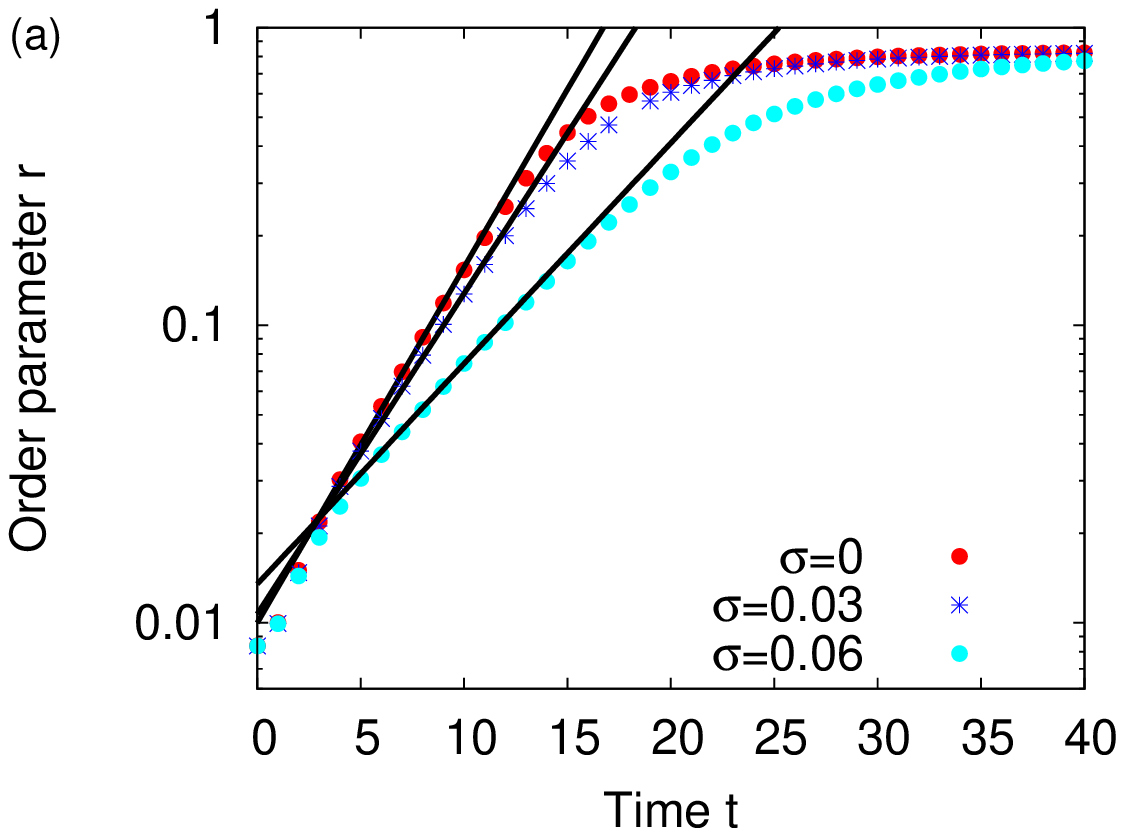}
\includegraphics[width=80mm]{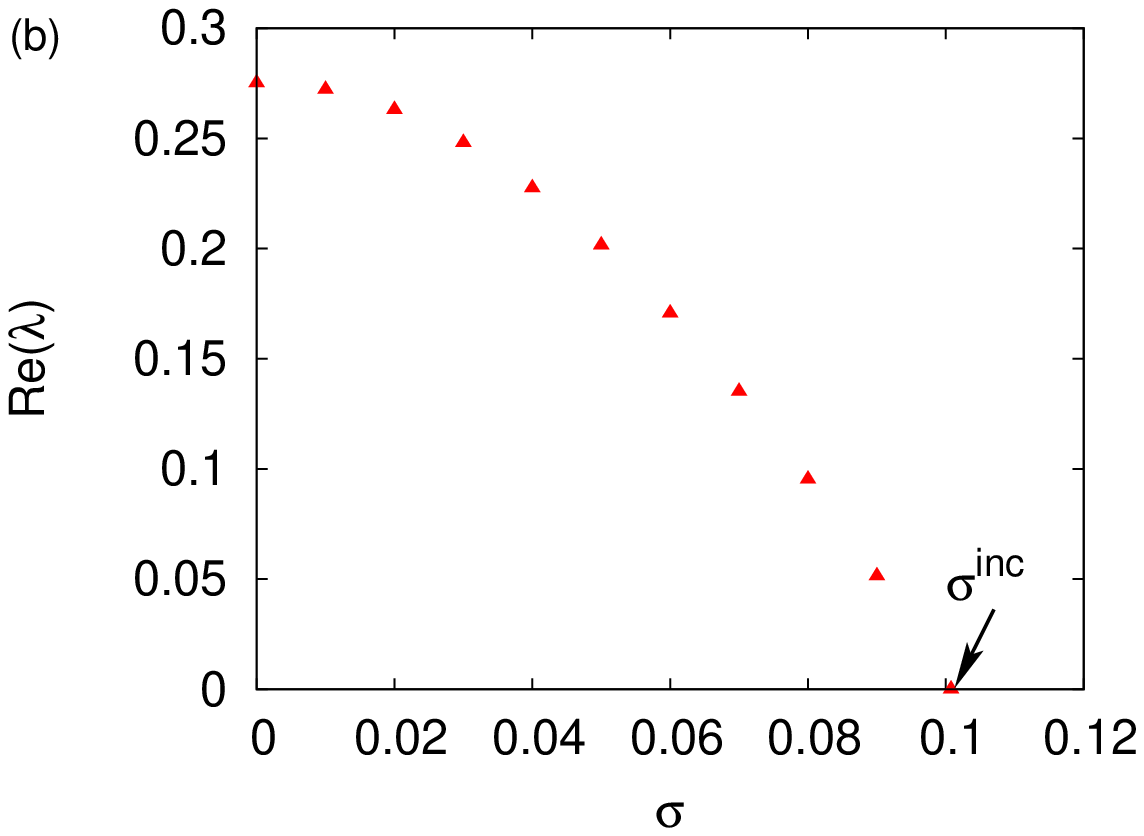}
\caption{(Color online) (a) Simulation results denoted by points,
demonstrating exponentially fast relaxation $\sim e^{\lambda t}$ of $r$ from
its initial incoherent state value to its final synchronized state value for
values of $\sigma$ below $\sigma^{\rm inc}(m,T)\approx 0.10076$ for a Gaussian $g(\omega)$
with $m=20,T=0.25,N=10^4$; the black solid lines denote exponential
growth with theoretically computed growth rates $\lambda$ obtained from Eq. (\ref{stability-eqn}) for a
Gaussian $g(\omega)$ with zero mean and unit width. The simulation data are
obtained from $N$-body simulation for a Gaussian $g(\omega)$ with zero
mean and unit width. (b) Theoretical $\lambda$ as a function of
$\sigma$ for the same $m$ and $T$ values; in particular, $\lambda$ hits
zero at the stability threshold $\sigma^{\rm inc}(m,T)$.} 
\l{fig7}
\end{figure}

\begin{figure}[h!]
\centering
\includegraphics[width=80mm]{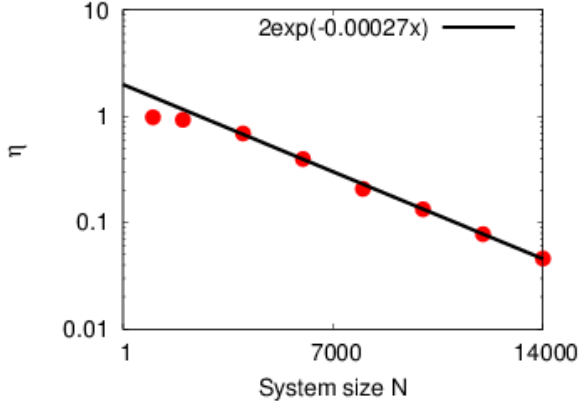}
\caption{(Color online) For $m=20,T=0.25,\sigma=0.11$, the figure shows the fraction $\eta$ of realizations of initial incoherent state
relaxing to synchronized state within the fixed time of observation $t=200$,
for a value of $\sigma$ above $\sigma^{\rm inc}(m,T)$, for which the
incoherent phase is linearly stable in the continuum limit. The figure
shows that $\eta$ for large $N$ decreases exponentially fast with
increase of $N$. The data are obtained in $N$-body simulations for a
Gaussian $g(\omega)$ with zero mean and unit width.} 
\l{fig8}
\end{figure}

\section{Conclusions}
To summarize, we considered an extension of the Kuramoto model
that includes an inertial term and a stochastic noise. For a general
unimodal frequency distribution, we obtained the complete phase diagram of
the model, demarcating parameter ranges to observe synchronization. We
showed that the system displays a nonequilibrium first-order transition
from a synchronized phase at low parameter values to an incoherent phase
at high values. The phase diagram contains all previous results derived
in specific limits of the dynamics. While we provided strong
numerical evidence for the existence of both the synchronized and the
incoherent phase, only the latter could be treated analytically to
obtain the corresponding linear stability threshold that bounds the
first-order transition point from below.
It would be interesting to consider possible extension of our studies to
systems with non-mean-field couplings, taking hints from similar previous
studies in specific limits of the dynamics
\cite{Campa:2003,Bachelard:2011,Gupta:2012,Gupta:2012-1}.

We acknowledge fruitful discussions with T. Dauxois, D. Mukamel, C. Nardini, A.
Patelli and H. Touchette, support of ENS-Lyon and the grants CEFIPRA 4604-3 and ANR-10-CEXC-010-01. 

\appendix
\section{Proof that the dynamics Eq. (\ref{eom-scaled}) does not satisfy
detailed balance}
\l{app0}
In this section, we prove that the dynamics Eq. (\ref{eom-scaled}) does not satisfy
detailed balance unless $g(\omega)=\delta(\omega)$, thus $\sigma$ is zero. 
For simplicity, we discuss the proof here for the case of two
distinct natural frequencies [bimodal $g(\omega)$].
Let us say that in a given realization of $g(\omega)$, there are $N_{1}$
oscillators with natural frequencies $\omega_{1}$, and $N_{2}$ oscillators with
natural frequencies $\omega_{2}$, where $N_{1}+N_{2}=N$.

To prove that the dynamics Eq. (\ref{eom-scaled}) does not satisfy detailed balance
unless $\sigma=0$, we rewrite the Fokker-Planck Eq. (\ref{eq:fp-eqn}) as
\begin{eqnarray}
&&\frac{\partial f_{N}(\mathbf{x})}{\partial t}=-\sum_{i=1}^{2N}\frac{\partial(A_{i}(\mathbf{x})f_{N}(\mathbf{x}))}{\partial
x_{i}}\nonumber \\
&&+\frac{1}{2}\sum_{i,j=1}^{2N}\frac{\partial^{2}(B_{i,j}(\mathbf{x})f_{N}(\mathbf{x}))}{\partial x_{i}\partial x_{j}},\label{eq:fp-compact}
\end{eqnarray}
where 
\begin{equation}
x_i=\left\{ 
\begin{array}{ll}
               \theta_{i};i=1,2,\dots,N,  \\
               v_{i-N};i=N+1,\dots,2N,  
               \end{array}
        \right. \\ 
               \label{eq:x-defn} 
\end{equation}
and  
\begin{eqnarray}
\mathbf{x} & =\{x_{i}\}_{1\le i\le2N}.\label{eq:boldx-defn}
\end{eqnarray}
In Eq. (\ref{eq:fp-compact}), the drift vector $A_{i}(\mathbf{x})$
is given by
\begin{equation}
A_{i}(\mathbf{x}) =\left\{
\begin{array}{ll}
                 v_{i};i=1,2,\dots,N, \\
                 -\frac{1}{\sqrt{m}}v_{i-N}
                 +\frac{1}{N}\sum_{j=1}^{N}\sin(\theta_{j}-\theta_{i-N})\\
                 +\sigma\Big(\Omega^{T}\Big)_{i-N};
                 i=N+1,\dots,2N,
                 \end{array}
           \right. \\
                 \label{eq:Ai-defn}
\end{equation}
while the diffusion matrix is 
\begin{equation}
B_{i,j}(\mathbf{x})  =\left\{
\begin{array}{ll}
              \frac{2T}{\sqrt{m}}\delta_{ij};i,j>N, \\
              0, ~{\rm Otherwise.}
              \end{array}
           \right. \\
           \label{eq:Bij-defn}
\end{equation}

The dynamics described by the Fokker-Planck equation of the form Eq. (\ref{eq:fp-compact})
satisfies detailed balance if and only if the following conditions
are satisfied \cite{Gardiner:1983}:
\begin{eqnarray}
&&\epsilon_{i}\epsilon_{j}B_{i,j}(\epsilon\mathbf{x}) =B_{i,j}(\mathbf{x}),\label{eq:detailed-balance-cond1}\\
&&\epsilon_{i}A_{i}(\epsilon\mathbf{x})f_{N}^{s}(\mathbf{x})
=-A_{i}(\mathbf{x})f_{N}^{s}(\mathbf{x})+\sum_{j=1}^{2N}\frac{\partial
B_{i,j}(\mathbf{x})f_{N}^{s}(\mathbf{x})}{\partial x_{j}},\nonumber \\ \label{eq:detailed-balance-cond2}
\end{eqnarray}
where $f_{N}^{s}(\mathbf{x})$ is the stationary solution of Eq. (\ref{eq:fp-compact}).
Here, $\epsilon_{i}=\pm1$ is a constant that denotes the parity with
respect to time reversal of the variables $x_{i}$s: Under time reversal,
the latter transform as $x_{i} \rightarrow\epsilon_{i}x_{i}$, where $\epsilon_{i}=-1$ or $+1$ depending on whether $x_{i}$ is
odd or even under time reversal. In our case, $\theta_{i}$s
are even, while $v_{i}$s are odd.

Using Eq. (\ref{eq:Bij-defn}), we see that the condition Eq. (\ref{eq:detailed-balance-cond1}) is trivially satisfied for our model. To check
the other condition, we formally solve Eq. (\ref{eq:detailed-balance-cond2})
for $f_{N}^{s}(\mathbf{x})$ and check if the solution solves Eq.
(\ref{eq:fp-compact}) in the stationary state. From Eq. (\ref{eq:detailed-balance-cond2}),
we see that for $i=1,2,\dots,N$, the condition reduces to 
\begin{eqnarray}
\epsilon_{i}A_{i}(\epsilon\mathbf{x})f_{N}^{s}(\mathbf{x}) & =-A_{i}(\mathbf{x})f_{N}^{s}(\mathbf{x}),
\end{eqnarray}
which, using Eq. (\ref{eq:Ai-defn}), is obviously satisfied. For $i=N+1,\dots,2N$,
we have 
\begin{eqnarray}
v_{k}f_{N}^{s}(\mathbf{x}) & =-\frac{T\partial f_{N}^{s}(\mathbf{x})}{\partial v_{k}};k=i-N.\label{eq:cond2-eq1}
\end{eqnarray}
Solving Eq. (\ref{eq:cond2-eq1}), we get 
\begin{eqnarray}
f_{N}^{s}(\mathbf{x}) & \propto
d(\theta_{1},\theta_{2},\dots,\theta_{N})\exp\Big[-\frac{1}{2T}\sum_{k=1}^{N}v_{k}^{2}\Big],
\nonumber \\
\label{eq:stationary-soln1}
\end{eqnarray}
where $d(\theta_{1},\theta_{2},\dots,\theta_{N})$ is a yet undetermined
function.
Substituting Eq. (\ref{eq:stationary-soln1}) into Eq. (\ref{eq:fp-compact}),
and requiring that it is a stationary solution, we get that $\sigma$
has to be equal to zero and that
$d(\theta_{1},\theta_{2},\dots,\theta_{N})=\exp\Big(-\frac{1}{2NT}\sum_{i,j=1}^{N}\Big[1-\cos(\theta_{i}-\theta_{j})\Big]\Big)$.
Thus, for $\sigma=0$, when the dynamics reduces to
that of the BMF model, we get the stationary
solution as
\begin{equation}
f_{N,\sigma=0}^{s}(\mathbf{z}) \propto\exp\Big[-\frac{H}{T}\Big].\label{eq:bmf-soln}
\end{equation}
where $H$ is the Hamiltonian (expressed in
terms of dimensionless variables introduced above). 
The lack of detailed balance for $\sigma \ne 0$ obviously extends to any distribution
$g(\omega)$.

\section{Simulation details} 
\l{app-simulation}
Here we describe the method to simulate the dynamics Eq. (\ref{eom-scaled}) for
given values of $m,T,\sigma$ (note that we are dropping overbars for
simplicity of notation), and for a given realization of
$\omega_i$'s, by employing a numerical integration scheme
\cite{Gupta:2012-2}. To simulate the dynamics over a time interval $[0:\mathcal{T}]$, we first choose a time step size
$\Delta t \ll 1$. Next, we set $t_n=n\Delta t$ as the $n$-th time step
of the dynamics, where $n=0,1,2,\ldots,N_t$, and
$N_t=\mathcal{T}/\Delta t$. In the numerical scheme, we first discard at
every time step the effect of the noise (i.e., consider
$1/\sqrt{m}=0$), and employ a fourth-order symplectic
algorithm to integrate the resulting symplectic part of the
dynamics \cite{McLachlan}. Following this, we add the effect of noise, and implement an Euler-like
first-order algorithm to update the dynamical variables.   
Specifically, one step of the scheme from $t_n$ to $t_{n+1}=t_n+\Delta t$ involves the following
updates of the dynamical variables for $i=1,2,\ldots,N$: For the symplectic part, we have, for $k=1,\ldots,4$, 
\begin{widetext}
\begin{eqnarray}
&&v_i\Big(t_{n}+\frac{k\Delta t}{4}\Big)=v_i\Big(t_n+\frac{(k-1)\Delta
t}{4}\Big)+b(k)\Delta t\Big[r\Big(t_n+\frac{(k-1)\Delta
t}{4}\Big)\sin\Big\{\psi\Big(t_n+\frac{(k-1)\Delta
t}{4}\Big)-\th_i\Big(t_n+\frac{(k-1)\Delta
t}{4}\Big)\Big\}+\sigma\omega_i\Big]; \nonumber \\
&&r\Big(t_n+\frac{(k-1)\Delta
t}{4}\Big)=\sqrt{r_x^2+r_y^2},~~\psi\Big(t_n+\frac{(k-1)\Delta
t}{4}\Big)=\tan^{-1}\frac{r_y}{r_x}, \nonumber \\
&&r_x=\frac{1}{N}\sum_{j=1}^N
\sin\Big[\th_j\Big(t_n+\frac{(k-1)\Delta t}{4}\Big)\Big],~~r_y=\frac{1}{N}\sum_{j=1}^N
\cos\Big[\th_j\Big(t_n+\frac{(k-1)\Delta t}{4}\Big)\Big], \nonumber \\
\label{formalintegration1} \\ 
&&\th_i\Big(t_{n}+\frac{k\Delta t}{4}\Big)=\th_i\Big(t_n+\frac{(k-1)\Delta
t}{4}\Big)+a(k)\Delta t ~v_i\Big(t_n+\frac{k\Delta t}{4}\Big),
\label{formalintegration2} 
\end{eqnarray}
where the constants $a(k)$'s and $b(k)$'s are obtained from Ref.
\cite{McLachlan}: one has 
\bea
&& a(1)=0.5153528374311229364, ~~a(2)=-0.085782019412973646, \nonumber \\
&&a(3)=0.4415830236164665242, ~~a(4)=0.1288461583653841854, \nonumber \\ 
&&b(1)=0.1344961992774310892, ~~b(2)=-0.2248198030794208058, \nonumber \\
&&b(3)=0.7563200005156682911, ~~b(4)=0.3340036032863214255.
\eea
\end{widetext}
At the end of the updates Eqs. (\ref{formalintegration1}) and
(\ref{formalintegration2}), we have the set
$\{\th_i(t_{n+1}),v_i(t_{n+1})\}$. Next, we include the effect of the
stochastic noise by keeping $\th_i(t_{n+1})$'s unchanged,  but by
updating $v_i(t_{n+1})$'s as
\begin{equation}
v_i(t_{n+1}) \to v_i(t_{n+1})\Big[1-\frac{1}{\sqrt{m}} \Delta
t\Big]+\sqrt{2\Delta t\frac{T}{\sqrt{m}}}\Delta X(t_{n+1}).
\label{formalintegration3}
\end{equation}
Here $\Delta X$ is a  Gaussian distributed
random number with zero mean and unit variance.


\begin{thebibliography}{99}
\bibitem{width}We take the width of a Gaussian distribution to refer to
its standard deviation.
\bibitem{footnote}For non-zero $\sigma$, one should
draw landscapes for the free-energy-like large
deviation functional; we assume here the landscape picture of
phase transitions to also hold for this quantity.
\end{thebibliography}
\end{document}